\documentclass{emulateapj}

\usepackage{xcolor}
\usepackage{hyperref}
\usepackage{breakurl}
\usepackage{amsmath}
\usepackage{graphicx}
\usepackage{verbatim}
\usepackage{booktabs}
\usepackage{tabularx}
\usepackage[flushleft]{threeparttable}
\usepackage{ccaption}
\usepackage{color}

\newcommand{\be}{\begin{equation}}
\newcommand{\ee}{\end{equation}}

\newcommand{\rsun}{R$_\odot$}

\newcommand{\mearth}{M$_\oplus$}
\newcommand{\rearth}{R$_\oplus$}
\newcommand{\fearth}{F$_\oplus$}
\newcommand{\kms}{\ensuremath{\rm km\,s^{-1}}}

\newcommand{\thisstar}{Kepler-1655}
\newcommand{\thisplanet}{Kepler-1655b}

\newcommand{\prot}{13.6}
\newcommand{\prote}{1.4}
\newcommand{\tev}{23}
\newcommand{\teve}{8}
\newcommand{\kamp}{1.47}
\newcommand{\kampeu}{0.88}
\newcommand{\kampel}{0.80}

\newcommand{\mupp}{10.1}
\newcommand{\rhopl}{2.5}
\newcommand{\rhopleu}{1.6}
\newcommand{\rhoplel}{1.4}
\newcommand{\flux}{155}
\newcommand{\fluxe}{7}
\newcommand{\mass}{5.0}
\newcommand{\masseu}{3.1}
\newcommand{\massel}{2.8}

\newcommand{\ldone}{0.403}
\newcommand{\uldone}{0.077}
\newcommand{\ldtwo}{0.260}
\newcommand{\uldtwo}{0.039}
\newcommand{\rprstb}{0.01965}
\newcommand{\urprstb}{0.00069}
\newcommand{\arstb}{20.5}
\newcommand{\uarstb}{4.1}
\newcommand{\inclb}{87.62}
\newcommand{\uinclb}{0.55}
\newcommand{\impb}{0.85}
\newcommand{\uimpb}{0.13}
\newcommand{\rplb}{2.213}
\newcommand{\urplb}{0.082}
\newcommand{\perplb}{11.8728787}
\newcommand{\uperplb}{0.0000085}
\newcommand{\ttransitb}{2455013.89795}
\newcommand{\uttransitb}{0.00069}

\slugcomment{}

\shorttitle{\thisplanet}
\shortauthors{Haywood R. D. et al.}

\begin{document}


\title{An accurate mass determination for \thisplanet, a moderately-irradiated world with a significant volatile envelope}
\author{Rapha\"{e}lle D. Haywood\altaffilmark{1 $\dagger$},
Andrew Vanderburg\altaffilmark{1,2 $\dagger$},  
Annelies Mortier\altaffilmark{3},
Helen A. C. Giles\altaffilmark{4}, 
Mercedes L\'{o}pez-Morales\altaffilmark{1},
Eric D. Lopez\altaffilmark{5}, 
Luca Malavolta\altaffilmark{6,7},
David Charbonneau\altaffilmark{1},
Andrew Collier Cameron\altaffilmark{3},
Jeffrey L. Coughlin\altaffilmark{8},
Courtney D. Dressing\altaffilmark{9,10 $\dagger$},
Chantanelle Nava\altaffilmark{1},
David W. Latham\altaffilmark{1}, 
Xavier Dumusque\altaffilmark{4},
Christophe Lovis\altaffilmark{4}, 
Emilio Molinari\altaffilmark{11,12},
Francesco Pepe\altaffilmark{4}, 
Alessandro Sozzetti\altaffilmark{13},
St\'ephane Udry\altaffilmark{4},
Fran\c{c}ois Bouchy\altaffilmark{4},
John A. Johnson\altaffilmark{1},
Michel Mayor\altaffilmark{4},
Giusi Micela\altaffilmark{14},
David Phillips\altaffilmark{1},
Giampaolo Piotto\altaffilmark{6,7},
Ken Rice\altaffilmark{15,16},
Dimitar Sasselov\altaffilmark{1},
Damien S\'egransan\altaffilmark{4},
Chris Watson\altaffilmark{17},
Laura Affer\altaffilmark{14},  
Aldo S. Bonomo\altaffilmark{13},
Lars A. Buchhave\altaffilmark{18},
David R. Ciardi\altaffilmark{19},
Aldo F. Fiorenzano\altaffilmark{11}
and Avet Harutyunyan\altaffilmark{11}
}

\email{rhaywood@cfa.harvard.edu}
\altaffiltext{$\dagger$}{NASA Sagan Fellow}
\altaffiltext{1}{Harvard-Smithsonian Center for Astrophysics, 60 Garden Street, Cambridge, MA 01238, USA}
\altaffiltext{2}{Department of Astronomy, The University of Texas at Austin, 2515 Speedway, Stop C1400, Austin, TX 78712, USA}
\altaffiltext{3}{Centre for Exoplanet Science, SUPA, School of Physics and Astronomy, University of St Andrews, St Andrews KY16 9SS, UK}
\altaffiltext{4}{Observatoire Astronomique de l'Universit\'e de Gen\`eve, Chemin des Maillettes 51, Sauverny, CH-1290, Switzerland}
\altaffiltext{5}{NASA Goddard Space Flight Center, 8800 Greenbelt Road, Greenbelt, MD 20771, USA}
\altaffiltext{6}{Dipartimento di Fisica e Astronomia ``Galileo Galilei", Universita' di Padova, Vicolo dell'Osservatorio 3, I-35122 Padova, Italy}
\altaffiltext{7}{INAF - Osservatorio Astronomico di Padova, Vicolo dell'Osservatorio 5, 35122 Padova, Italy}
\altaffiltext{8}{SETI Institute, 189 Bernardo Avenue Suite 200, Mountain View, CA 94043, USA}
\altaffiltext{9}{Division of Geological \& Planetary Sciences, California Institute of Technology, Pasadena, CA 91125}
\altaffiltext{10}{Astronomy Department, University of California, Berkeley, CA 94720, USA}
\altaffiltext{11}{INAF - Fundaci\'on Galileo Galilei, Rambla Jos\'e Ana Fernandez P\'erez 7, E-38712 Bre$\tilde{\rm n}$a Baja, Tenerife, Spain}
\altaffiltext{12}{INAF - Osservatorio Astronomico di Cagliari, via della Scienza 5, 09047, Selargius, Italy}
\altaffiltext{13}{INAF - Osservatorio Astrofisico di Torino, via Osservatorio 20, 10025 Pino Torinese, Italy}
\altaffiltext{14}{INAF - Osservatorio Astronomico di Palermo, Piazza del Parlamento 1, 90134 Palermo, Italy}
\altaffiltext{15}{SUPA, Institute for Astronomy, Royal Observatory, University of Edinburgh, Blackford Hill, Edinburgh EH93HJ, UK}
\altaffiltext{16}{Centre for Exoplanet Science,  University of Edinburgh,  Edinburgh,  UK}
\altaffiltext{17}{Astrophysics Research Centre, School of Mathematics and Physics, Queen's University Belfast, Belfast, BT7 1NN, UK}
\altaffiltext{18}{Centre for Star and Planet Formation, Natural History Museum of Denmark, University of Copenhagen, DK-1350 Copenhagen, Denmark}
\altaffiltext{19}{NASA Exoplanet Science Institute, Caltech/IPAC-NExScI, 1200 East California Blvd, Pasadena, CA 91125, USA}

%




\begin{abstract}
We present the confirmation of a small, moderately-irradiated ($F=155\pm7$ \fearth) Neptune with a substantial gas envelope in a $P$=\perplb$\pm$\uperplb-day orbit about a quiet, Sun-like G0V star \thisstar. Based on our analysis of the \emph{Kepler} light curve, we determined \thisplanet's radius to be \rplb$\pm$\urplb \,\rearth. We acquired 95 high-resolution spectra with TNG/HARPS-N, enabling us to characterize the host star and determine an accurate mass for \thisplanet\, of \mass$\pm^{\masseu}_{\massel}$ \,\mearth \, via Gaussian-process regression. Our mass determination excludes an Earth-like composition with 98\% confidence.
\thisplanet \ falls on the upper edge of the evaporation valley, in the relatively sparsely occupied transition region between rocky and gas-rich planets. It is therefore part of a population of planets that we should actively seek to characterize further.

\end{abstract}


\section{Introduction}

In our own solar system, we see a sharp transition between the inner planets, which are small ($R_p \leq$ 1 \rearth) and rocky, and the outer planets that are larger ($R_p \geq 3.88$ \rearth), much more massive, and have thick, gaseous envelopes. For exoplanets with radii intermediate to that of the Earth (1 \rearth) and Neptune (3.88 \rearth), several factors go into determining whether planets acquire or retain a thick gaseous envelope. Several studies have determined statistically from radius and mass determinations of exoplanets that most planets smaller than 1.6 \rearth\ are rocky, \emph{i.e.} they do not  have large envelopes but only a thin, secondary atmosphere, if any at all  \citep{Rogers2015, weissmarcy,Dressing2015,LopezRice2016,Lopez2016,LopezFortney2014,2016AJ....152..160B,Gettel2016}. Others have found that planets in less irradiated orbits tend to be more likely to have gaseous envelopes than more highly irradiated planets \citep{2014Hadden, JontofHutter2016}. However, it is still unclear under which circumstances a planet will obtain and retain a thick gaseous envelope and how this is related to other parameters, such as stellar irradiation levels.

The characterization of the mass of a small planet in an orbit of a few days to a few months around a Sun-like star (\emph{i.e.} in the incident flux range $\approx$ 1-5000 \fearth) is primarily limited by the stellar magnetic features acting over this timescale and producing RV variations that compromise our mass determinations. Magnetic fields produce large, dark starspots and bright faculae on the stellar photosphere. These features induce RV variations modulated by the rotation of the star and varying in amplitude as the features emerge, grow and decay. There are two physical processes at play: (i) dark starspots and bright faculae break the Doppler balance between the approaching blueshifted stellar hemisphere and the receding redshifted half of the star \citep{1997ApJ...485..319S,Lagrange2010I,Boisse:2012ju,2016MNRAS.457.3637H}; (ii) they inhibit the star's convective motions, and this suppresses part of the blueshift that naturally arises from convection \citep{Dravins:1981tl,2010A&A...512A..39M,2010A&A...519A..66M,2014ApJ...796..132D,2016MNRAS.457.3637H}. 

In this paper we report the confirmation of \thisplanet, a mini Neptune orbiting a Sun-like star, first noted as a planet candidate (KOI-280.01) by \cite{koi2}.
\thisplanet\, straddles the valley between the small, rocky worlds and the larger, gas-rich worlds. It is also in a moderately-irradiated orbit.
We present the $Kepler$ and HARPS-N observations for this system in Section~\ref{observations}. Based on these datasets, we determine the properties of the host star (Section~\ref{analysis}), statistically validate \thisplanet\ as a planet (Section~\ref{validation}), measure \thisplanet's radius (Section~\ref{subsectrans}) and mass (Section~\ref{subsecrv}). Using these newly-determined stellar and planetary parameters, we place \thisplanet\ among other exoplanets found to date and  investigate the influence of incident flux on planets with thick gaseous envelopes, as compared with gas-poor, rocky planets (Section~\ref{discussion}). 

\section{Observations}\label{observations}

\subsection{Kepler Photometry} \label{kepler}

\thisstar\ was monitored with {\it Kepler} in 29.4 min, long-cadence mode between quarters Q0 and Q17, and in 58.9 sec, short-cadence mode in quarters Q2-Q3 and Q6-Q17, covering a total time period of 1,459.49 days (BJD 2454964.51289 -- 2456424.00183). 

The simple aperture flux (SAP) shows large long-term variations on the timescale of a \emph{Kepler} quarter due to differential velocity aberration, which without adequate removal obscures astrophysical stellar rotation signals as small as those expected for \thisstar. The Presearch Data Conditioning (PDC) reduction from Data Release 25 (DR25) did not remove these long-term trends completely due to an inadequate choice of aperture pixels. The PDC reduction from DR21, however, had a particular choice of apertures which was much more effective at removing these trends.
We therefore worked with the PDCSAP light curve from Data Release 21 \citep{2014PASP..126..100S,2012PASP..124.1000S,2012PASP..124..985S} to estimate the stellar and planet parameters. 

\begin{figure*}[t]
\centering
\includegraphics[scale=0.57]{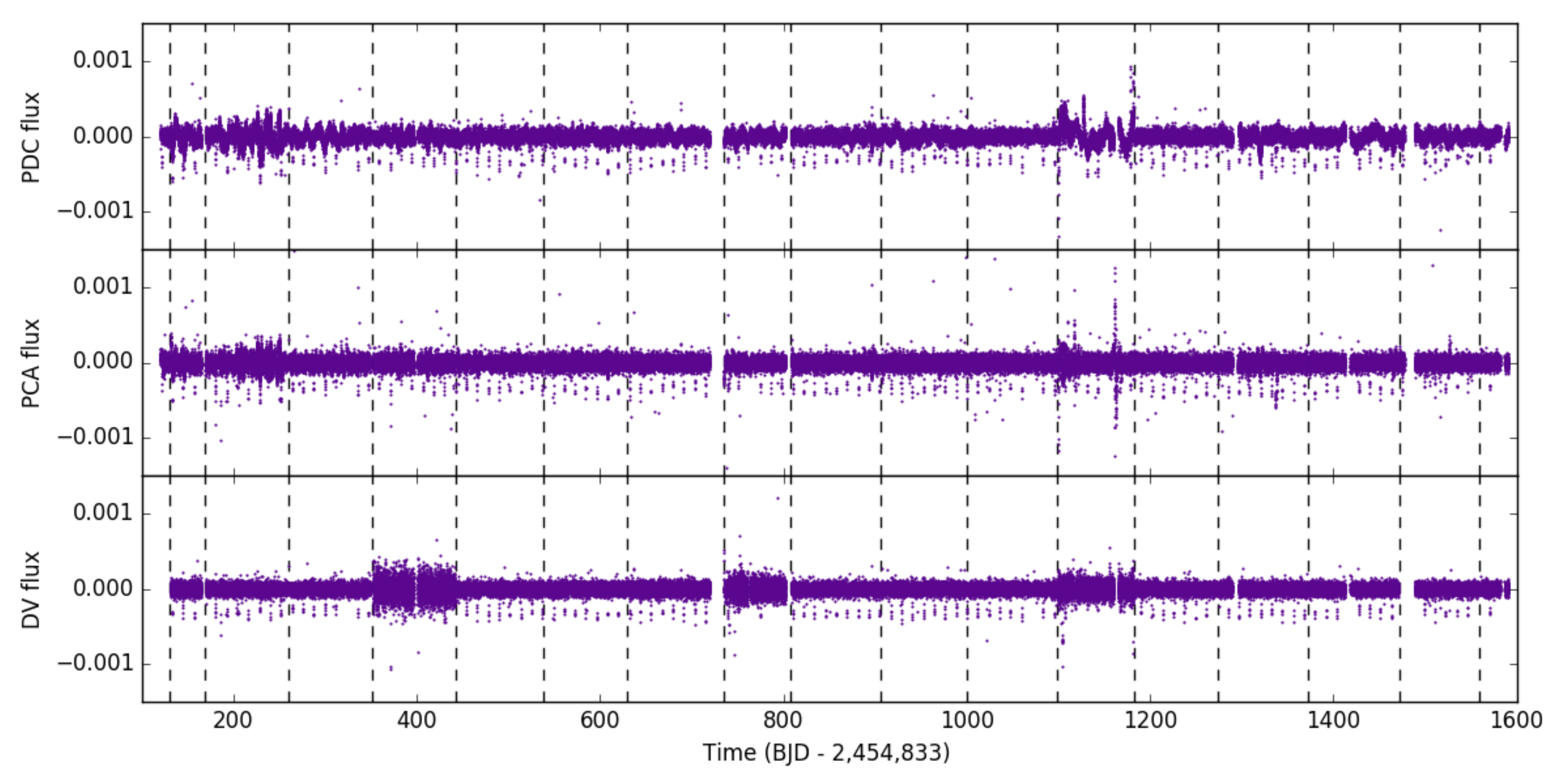}
\caption{Full Q0--Q17 long cadence \emph{Kepler} lightcurve detrended using: {\it Top} -- Presearch Data Conditioning Simple Aperture Photometry (PDCSAP, DR21); {\it Middle} -- Principal Component Analysis (PCA); {\it Bottom} -- Data Validation (DV). The dashed lines mark the start of each \emph{Kepler} quarter. \label{lcs}}
\end{figure*}

We compared the PDC (DR21) light curve with the principle component analysis (PCA) light curve and the Data Validation (DV) light curve, generated as described in \cite{2012AJ....143...39C}; see also \cite{LopezMorales2016} for a detailed description of these two types of analyses. 
All three lightcurves are plotted in Figure~\ref{lcs}.
The PDCSAP (DR 21) and PCA light curves show very similar features. They both display little variability aside from the transits of \thisplanet, which indicates that \thisstar~ is a quiet, low-activity star. 
Some larger dispersion is visible in quarters Q0-Q2, which is likely to be the signature of rotation-modulated activity (more on this in Section~\ref{arlifetime}). 
We note that Q12 has increased systematics in all three detrendings, possibly due to the presence of three coronal mass ejections which affected spacecraft and detector performance throughout the quarter \citep{2016ksci.rept....2B}. The DV detrending also shows increased systematics, most likely due to the harmonic removal module in DV, which operates on a per-quarter basis \citep{2017ksci.rept...11L}.

\subsection{HARPS-N Spectroscopy}
We observed \thisstar\ with the HARPS-N instrument \citep{harpsn} on the Telescopio Nazionale Galileo (TNG) at La Palma, Spain over two seasons between 2015 June 7 and 2016 November 13. 
The spectra were processed using the HARPS Data Reduction System (DRS)\citep{Baranne:1996tb}. The cross-correlation was performed using a G2 spectral mask \citep{2002Msngr.110....9P}. The RV measurements and the spectroscopic activity indicators are provided in Table~\ref{specdata}.
The median, minimum and maximum signal to noise ratio of the HARPS spectra at the centre of the spectral order number 50 are 51.8, 24.8 and 79.2, respectively.

The host star is fainter than typical RV targets and its RVs can be potentially affected by moonlight contamination. We followed the procedure detailed in \cite{2017AJ....153..224M} and determined that none of our measurements were affected, including those carried out near full Moon. In all cases the RV of the star with respect to the observer rest frame, \emph{i.e.} the difference between the systemic RV of the star and the barycentric RV correction, was higher than -25 km.s$^{-1}$, that is around three times the FWHM of the CCF, thus avoiding any moonlight contamination.

\section{Stellar properties of \thisstar}\label{analysis}
\label{sec:stellar_prop}

\thisstar\ is a G0V star with an apparent V magnitude of $11.05 \pm 0.08$, located at a distance of $230.41 \pm 28.14$\,pc from the Sun, according to the \emph{Gaia} data release DR1 \citep{Gaia2016}. All relevant stellar parameters can be found in Table \ref{tab:star}.

We added all individual HARPS-N spectra together and performed a spectroscopic line analysis. Equivalent widths of a list of iron lines (\ion{Fe}{1} and \ion{Fe}{2}) \citep{Sousa2011a} were automatically determined using ARESv2 \citep{Sousa2015}. We then used them, along with a grid of ATLAS plane-parallel model atmospheres \citep{Kurucz1993}, to determine the atmospheric parameters, assuming local thermodynamic equilibrium in the 2014 version of the MOOG code\footnote{\url{http://www.as.utexas.edu/~chris/moog.html}} \citep{Sneden2012}. We used the iron abundance as a proxy for the metallicity. More details on the method are found in \citet{Sousa2014} and references therein. We corrected the surface gravity resulting from this analysis to a more accurate value following \citet{Mortier2014}.

We quadratically added systematic errors to our precision errors, intrinsic to our spectroscopic method. For the effective temperature we added a systematic error of $60$\,K, for the surface gravity $0.1$\,dex, and for metallicity $0.04$\,dex \citep{Sousa2011a}.

We found an effective temperature of 6148\,K and a metallicity of -0.24. These values are consistent with the values reported by \citet{Huber2013} (6134\,K and -0.24, respectively), based on a spectral synthesis analysis of a TRES spectrum. 

As a sanity check we also estimated the temperature and metallicity from the HARPS-N CCFs according to the method of \cite{2017MNRAS.469.3965M} \footnote{\url{https://github.com/LucaMalavolta/CCFpams}} and obtained a similar result (6151 $\pm 34$ K, $-0.27 \pm 0.03$, internal errors only).

The stellar mass and radius were derived using a Bayesian estimation \citep{Dasilva2006} and a set of PARSEC isochrones \citep{Bressan2012}\footnote{http://stev.oapd.inaf.it/cgi-bin/param}. We used the effective temperature and metallicity from the spectroscopic analysis as input. We ran the analysis twice, once using the apparent V magnitude and parallax and once using the asteroseismic values $\Delta\nu$ and $\nu_{max}$ obtained by \citet{Huber2013}. The values are consistent, with the ones resulting from the asteroseismology being more precise. We use the latter throughout the rest of the paper (see Table~\ref{bigtable}). These mass and radius values are also consistent with the ones obtained by \citet{Huber2013} and \citet{SilvaAguirre2015}. The resulting stellar density is consistent with what is found by analysing the transit shape (see Section \ref{subsectrans}). This analysis also determined an age of $2.56\pm1.06$\,Gyr, consistent with the $3.27 \pm ^{0.59}_{0.64}$ Gyr from the analysis of \citet{SilvaAguirre2015}.

The spectral synthesis used by \citet{Huber2013} revealed a $v\,\sin{i_{\rm star}}$ of $3.5 \pm 0.5$\kms, making \thisstar\ a relatively slowly rotating star. In an asteroseismology analysis, \cite{campante} determined the stellar inclination to be between 38.4 and 90 degrees (within the 95.4\% highest posterior density credible region). 
This value translates into an upper limit for the rotation period of $14.8 \pm 2.4$ days, and a lower limit of $9.2 \pm 2.4$ days which are consistent with the rotation period we determine from the \emph{Kepler} light curve (see Section~\ref{arlifetime}).

\section{Statistical Validation}\label{validation}

The detection of a spectroscopic orbit in phase with the photometric ephemeris through RV observations is the gold standard for proving that transit signals found in \emph{Kepler}\ data are genuine exoplanets. In the case of \thisplanet, however, we do not detect the planet's reflex motion at high significance through our HARPS-N RV observations (see Section~\ref{subsecrv}). Instead, in this section, we show that the transit signal is very likely a genuine exoplanet by calculating the astrophysical false positive probabilities using the open source tool \texttt{vespa} \citep{morton2012, morton2015}, and by interpreting additional observations that are not considered by the \texttt{vespa} software.

\paragraph{Assessement of false positive probabilities using Vespa}
\texttt{Vespa} calculates the likelihood that a transit signal is caused by a planet compared to the likelihood that the transit signal is caused by some other astrophysical phenomenon such as an eclipsing binary, either on the foreground star, or on another star in the photometric aperture. \texttt{Vespa} compares the shape of the observed transit to what would be expected for these different scenarios, and imposes priors based on the density of stars in the field, constraints on other stars in the aperture from high resolution imaging, limits on putative secondary eclipses, and differences in the depths of odd and even eclipses (to constrain scenarios where the signal is caused by an eclipsing binary with double the orbital period we find). We include as constraints two adaptive optics images acquired with the Palomar PHARO-AO system in J and K bands, downloaded from the \emph{Kepler}\ Community Follow-Up Program (CFOP) webpage. In the case of \thisstar, we also impose the constraint that we definitively rule out scenarios where \thisplanet\ is actually an eclipsing binary based on our HARPS-N RV observations, because we have a strong upper limit on the mass measurement requiring that any companion in a short period orbit be planetary.

Given these constraints, we find a false positive probability of $2\times10^{-3}$ for \thisplanet, which is considerably lower than the $10^{-2}$ threshold commonly used to validate \emph{Kepler}\ candidates \citep{rowemultiplanets, morton2016}. The dominant false positive scenario 

is that the \thisstar\ system is a hierarchical eclipsing binary, where a physically associated low-mass eclipsing binary system near to \thisstar\ is causing the transit signal.

\paragraph{Additional observational constraints}

We see no evidence for the existence of a companion star to \thisstar~ according to AO imaging (see previous paragraph). The maximum peak-to-peak RV variation observed by HARPS-N is well below 20 m.s$^{-1}$(see Section~\ref{subsecrv}). These two observational constraints entirely rule out a foreground eclipsing binary scenario. This drops the false positive probability by about a factor of 10 from the \texttt{vespa} estimate and thus places the false positive probability well below the threshold of 1\% that is typically used.

The \emph{Kepler}\ short cadence data, which did not go into the original \texttt{vespa} analysis, puts further constraints on these scenarios. The dominant scenario that arises from the \texttt{vespa} calculations is the hierarchical scenario. We show that this is entirely ruled out by our short cadence data.
With the short cadence photometry, we resolve transit ingress and egress, measuring the duration of ingress/egress, $t_{1,2}$, to be 10 $\pm$ 3 minutes, with ingress and egress each taking up 7\% $\pm$ 2\% of the total mid-ingress to mid-egress transit duration, $t_{1.5,3.5}$. 
The ratio between the transit ingress/egress time and the duration, $f = t_{1,2}/t_{1.5,3.5}$, is a measurement of the largest possible companion to star radius ratio, independent of the amount of blending in the light curve. If we assume that the transit is caused by a background object, the faintest background object that could cause the signal we see is only a factor of $f^{2}/ (R_p/R_\star)^2$ = 12 $\pm$ 6 times fainter than \thisstar. For a physically-associated star, this brightness difference corresponds to roughly a late K-dwarf, with stellar radius of about 0.7 \rsun. The largest physically-associated object which could cause the transit shape we see is therefore about $R_{\rm companion} \simeq $0.7~\rsun $\times f \simeq 6$ \rearth, and therefore of planetary size. 

The last plausible scenario that remains is that of a hierarchical planet. Even though we cannot rule it out, it is a very unlikely scenario.
The stringent limits on false positive scenarios from our \texttt{vespa} analysis, the lack of evidence for a companion star, the fact that small planets are considerably more common than large planets, and the fact that we have a tentative detection of the spectroscopic orbit of \thisplanet\ all give us the highest confidence that \thisplanet\ is in fact a genuine planet transiting \thisstar.

\section{Radius of \thisplanet\ from transit analysis}\label{subsectrans}


We fit the PDCSAP short cadence light curves produced by the \emph{Kepler}\ pipeline of \thisstar. We flatten the light curve by fitting second order polynomials 

to the out-of-transit light curves near transits, and dividing the best-fit polynomial from the light curve. The PDCSAP short cadence light curves have had some systematics removed, but there are still a considerable number of discrepant data points in the light curve, especially towards the end of the original \emph{Kepler}\ mission, when the second of four reaction wheels was close to failure. We exclude outliers from the phase-folded light curve by dividing it into bins of a few minutes. Within each of these bins, we then exclude 3-sigma outliers, although we find that a more conservative 5-sigma clipping does not change the resulting planet parameters significantly.

\begin{figure*}[t]
\centering
\includegraphics[scale=0.6]{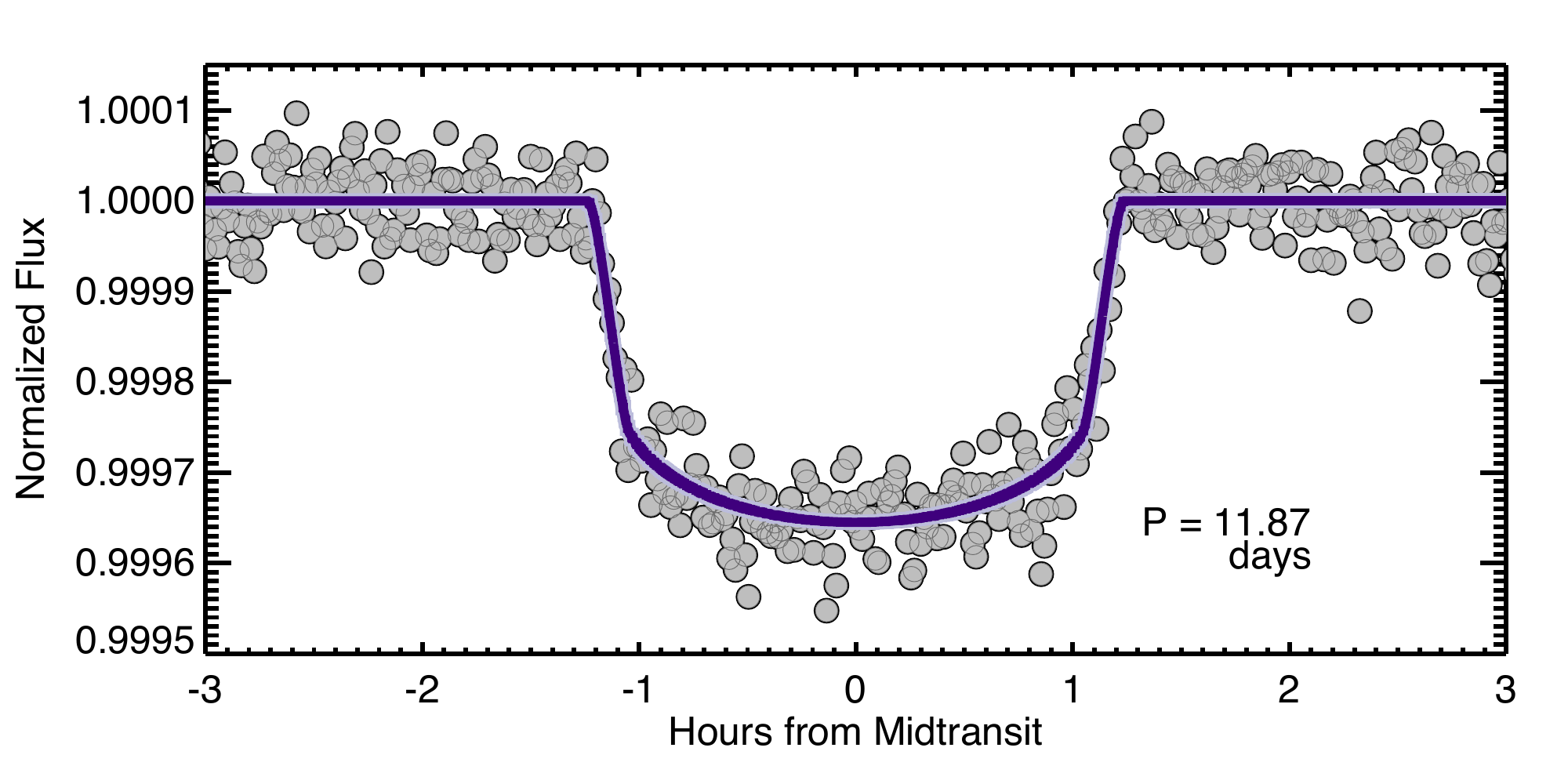}
\caption{Short-cadence \emph{Kepler}\ transit light curve of \thisplanet. Gray dots are the short cadence data binned in roughly 30-second intervals. The line is the maximum-likelihood transit model. \label{sclc}}
\end{figure*}

We then fit the transit light curve with a transit model \citep{mandelagol} using a Markov Chain Monte Carlo (MCMC) algorithm with an affine invariant sampler \citep{goodman}. We account for the 58.34-second short-cadence integration time by oversampling model light curves by a factor of 10 and performing a trapezoidal integration. We fit for the planetary orbital period, transit time, scaled semi-major axis ($a/R_\star$), the planetary to stellar radius ratio ($R_p/R_{\star}$), the orbital inclination, and quadratic limb darkening parameters $q_1$ and $q_2$, as defined by \citet{kippingld}. We impose Gaussian priors on the traditional limb darkening parameters $u_1$ and $u_2$, centered at the values predicted by \citet{Claret2011}, with widths of 0.07 in each parameter (which is the typical systematic uncertainty in model limb darkening parameters found by \citealt{mullerlimbdarkening}).  We sample the parameter space using an ensemble of 50 walkers, evolved for 20,000 steps. We confirm that the MCMC chains were well mixed by calculating the Gelman-Rubin convergence statistics \citep{gelmanrubin}.

A binned short-cadence transit light curve and the best-fit model is shown in Figure \ref{sclc}.

The ultra-precise \emph{Kepler}\ short cadence data resolves the transit ingress and egress for \thisplanet, and therefore is able to precisely measure the planetary impact parameter. We find that \thisplanet\ transits near the limb of its host star, with an impact parameter of 0.85~$^{+.03}_{-.07}$, which makes the radius ratio somewhat larger than would likely be inferred from a fit to the long-cadence data alone (without a prior placed on the stellar density and eccentricity).

\begin{figure}[t]
\centering
\includegraphics[scale=0.6]{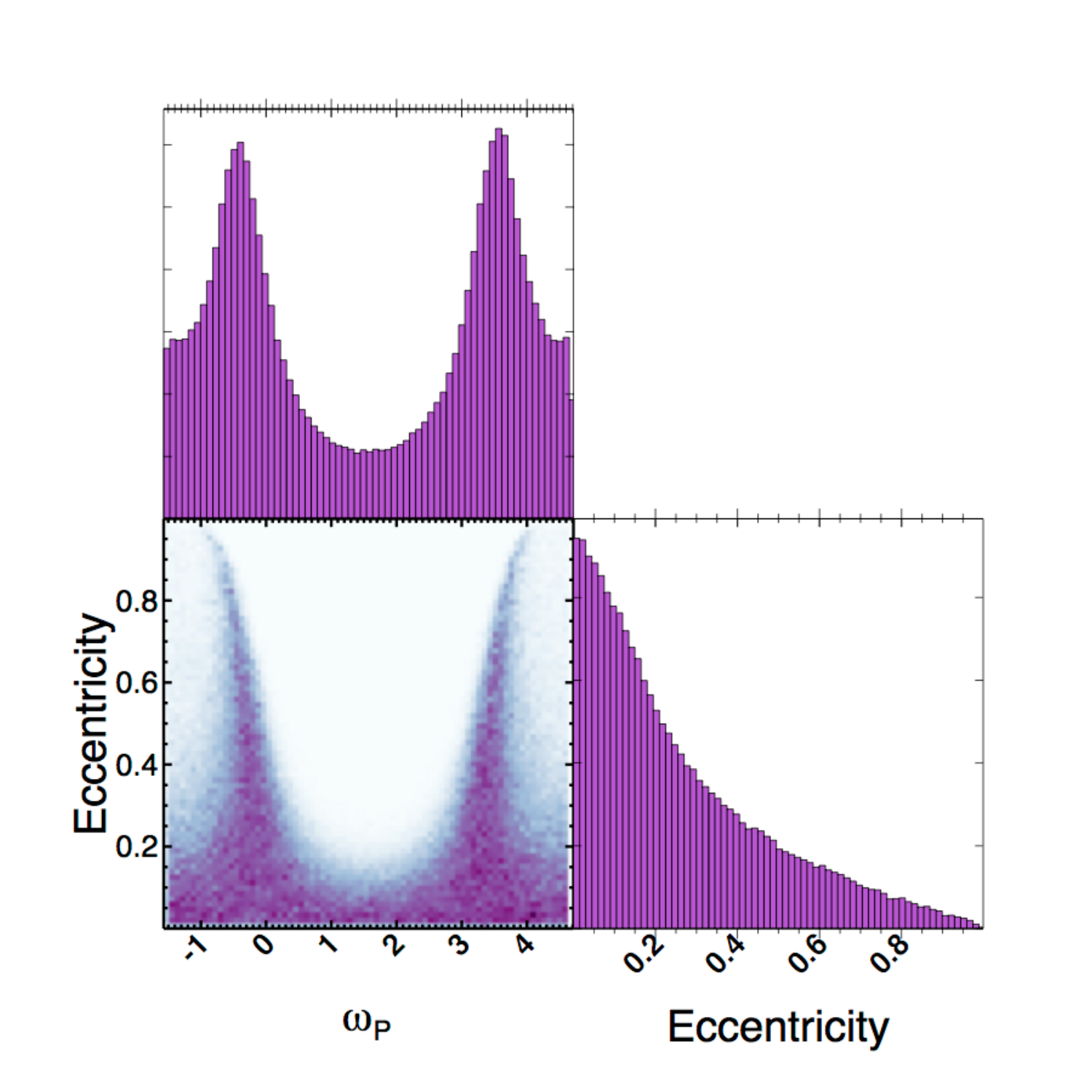}
\caption{Correlation plots between the orbital eccentricity of \thisplanet\ and its argument of periastron. Marginalized histograms of these parameters are shown alongside the correlation plot.  \label{eccwp}}
\end{figure}

As a sanity check, we also fit the transits of \thisplanet\ using the Data Validation (DV) long-cadence light curve (not including quarters Q4, Q8 and Q12) using EXOFAST-1 \citep{Eastman2013}. 

All parameter estimates fitted via this method are consistent with the results we obtained from our short-cadence analysis, including the eccentricity.

\begin{figure*}[t]

\centering
\includegraphics[scale=0.65]{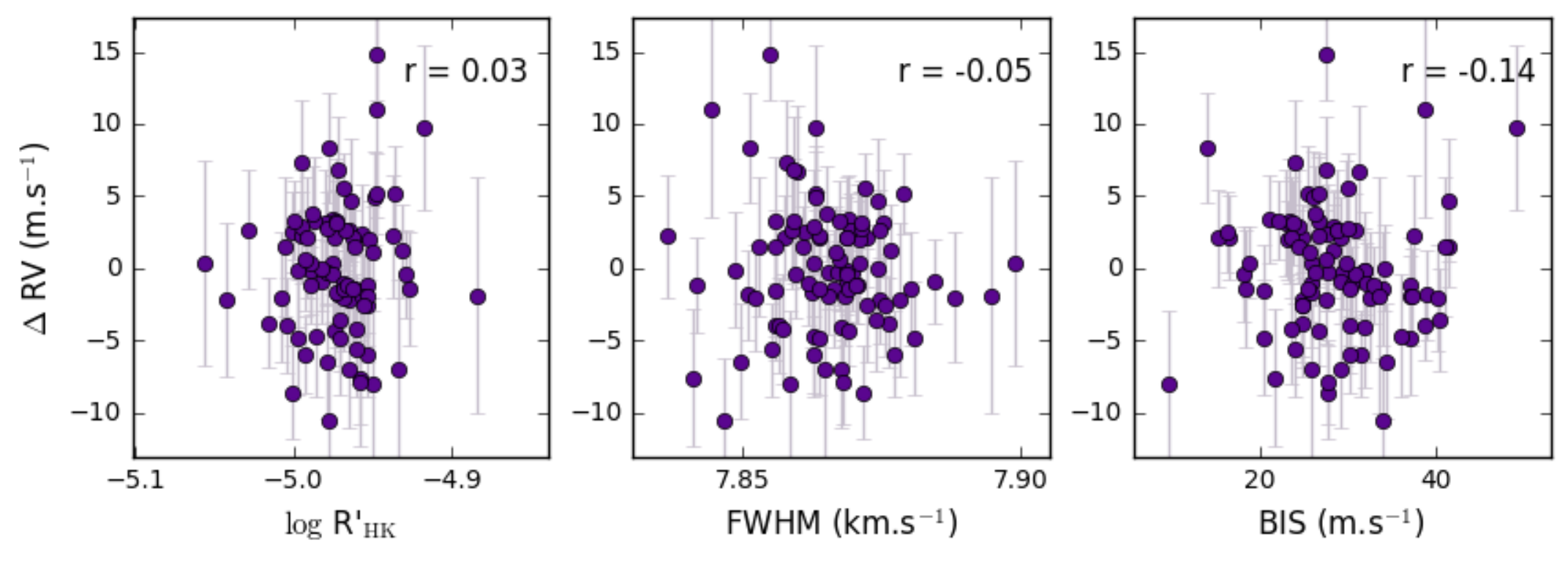}
\caption{Plots of the HARPS RV variations {\it vs.} the $\log R'_{\rm HK}$ index, the FWHM and the BIS of the cross-correlation function. The Spearman correlation coefficients for each pair of variables are given in the top right hand corner of each panel. We find no significant correlations. \label{fig:indicators}}
\end{figure*}

\subsection{Constraint on the eccentricity via asterodensity profiling}\label{asterodensity}

We placed constraints on the eccentricity, $e$, and argument of periastron, $\omega_p$, of \thisplanet's orbit by comparing our measured scaled semimajor axis ($a/R_\star$) from our short-cadence transit fits (see Section~\ref{subsectrans}) and the precisely-known asteroseismic stellar parameters (listed in Table~\ref{bigtable}). We followed the procedure outlined by \citet{dawsonjohnson} in their Section 3.4 and explored parameter space using an MCMC analysis with affine invariant ensemble sampling \citep{goodman}. We find that \thisplanet's orbit is consistent with circular, although some solutions with high eccentricity and finely tuned arguments of periastron are allowed. Our analysis gives 68\% and 95\% confidence upper limits of $e_{68\%}<0.31$ and $e_{95\%}<0.71$, respectively.  The two-dimensional probability distribution of allowed $e$ and $\omega_p$ is shown in Figure \ref{eccwp}. These distributions and upper limits are fully consistent with those obtained in our RV analysis (see Table~\ref{bigtable} and Figure~\ref{shark}).

\section{Mass of \thisplanet\ from RV analysis}\label{subsecrv}
The main obstacle to determining robust planet masses arises from the intrinsic magnetic activity of the host star.

\thisstar\ does not present particularly high levels of magnetic activity. In fact, the magnetic behavior exhibited in its light curve, spectroscopic activity indicators and RV curve is very similar to that of the Sun during its low-activity, ``quiet" phase.  
However, ongoing observations of the Sun as a star show activity-induced RV variations with an RMS of 1.6 m.s$^{-1}$ even though it is now entering the low phase of its 11-year magnetic activity cycle \citep{Dumusque:2015ApJ...814L..21D}. More generally, several large spectroscopic surveys have shown that even the quietest stars display activity-driven RV variations of order 1-2 m.s$^{-1}$ (\emph{eg.} the California Planet Search \citep{Isaacson:2010ud}; the HARPS-N Rocky Planet Search \citep{Motalebi2015}). 

In the current era of confirming and characterizing planets with reflex motions of 1-2 m.s$^{-1}$, accounting for the effects of magnetically-induced RV noise/signals, even in stars deemed to be ``quiet", becomes a necessary precaution. This is the only way we will determine planetary masses accurately and reliably (let alone precisely). 

In the case of \thisstar, we estimate that the rotationally-modulated, activity-induced RV variations have an RMS of order 0.5 m.s$^{-1}$. Furthermore, the stellar rotation and planetary orbital periods are very close to each other, at 13 and 11 days, respectively.
We perform an RV analysis based on Gaussian process (GP) regression, which can account for low-amplitude, quasi-periodic RV variations modulated by the star's rotation.

\subsection{Preliminary investigations}\label{activity}

Firstly, we perform some basic checks on the spectroscopic data available to us. We investigate whether the spectroscopically-derived activity indicators are reliable, and whether they provide any useful information for our analysis. Secondly, we determine the stellar rotation period and active-region evolution timescale from the PDCSAP lightcurve.
Thirdly, we look at the sampling strategy of the observations. In particular, we compare the two stellar timescales (rotation and evolution)  to the orbital period of \thisplanet \ and investigate how well all three timescales are sampled.

\begin{figure*}[t]
\centering
\includegraphics[scale=0.45]{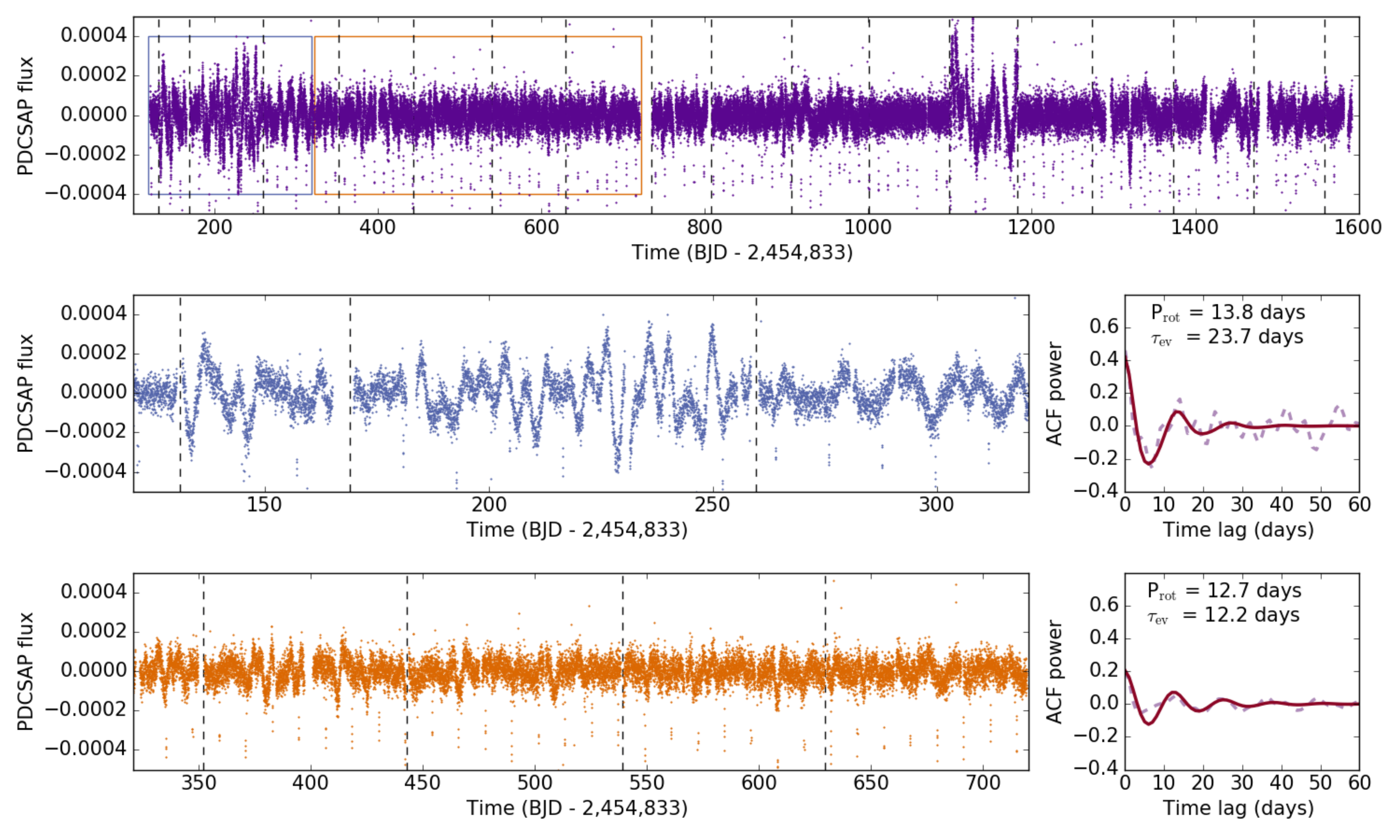}
\caption{Autocorrelation function (ACF) analysis. \emph{Top panel:} full PDCSAP light curve, including transits of \thisplanet. \emph{Middle panels:} zoom-in on a 200-day stretch of light curve during which the star is active, and corresponding ACF (dashed line), overlaid with our MCMC fit (solid line). \emph{Bottom panels:} zoom-in on a quiet 400-day stretch of the light curve, with corresponding ACF. Note that the transits were excluded for the computation of the ACFs. The dashed lines mark the start of each \emph{Kepler} quarter.
\label{fig:ACFplot}}
\end{figure*}

\subsubsection{``Traditional" spectroscopic activity indicators}

The average value of the $\log R'_{\rm HK}$ index (-4.97) is close to that of the Sun in its low-activity phase ($\approx -5.0$), implying that \thisstar\ is a relatively quiet star.

Figure~\ref{fig:indicators} shows the RV observations plotted against the ``traditional" spectroscopic activity indicators: the $\log R'_{\rm HK}$ index, computed from the DRS pipeline, which is a measure of the emission present in the core of the Ca {\sc ii} H \& K lines; the full width at half maximum (FWHM) and bisector span (BIS) of the cross-correlation function, which tell us about the asymmetry of the cross-correlation function \citep{Queloz:2001vs}. We see no significant correlations between the RVs and any of these activity indicators. This is expected as they are measurements that have been averaged over the whole stellar disc and small-scale structures such as spots and faculae, if present, are therefore likely to blur out. Moreover, the cross-correlation function is made up of many thousands of spectral lines whose shapes are all affected by stellar activity in different ways (depending on factors such as their formation depth, Land\'{e} factor, excitation potential, \emph{etc.}).

 \paragraph{Reliability of the $\log R'_{\rm HK}$ index for this star}
A recent study by \cite{Fossati2017} found that for stars further than about 100 pc, the Ca {\sc ii} H \& K line cores may be significantly affected by absorption from the interstellar medium (ISM), if the velocity of the ISM is close to that of the star, and the column density in the ISM cloud is high. 
This ISM-induced effect lowers the value of the $\log R'_{\rm HK}$ index, making the stars look less active than they really are. 
Although \cite{Fossati2017} note that this effect should be stable  over a timescale of years (even decades), they do caution us that it can mask the variability in the core of the Ca {\sc ii} H \& K lines and thus compromise the reliability of the $\log R'_{\rm HK}$ index as an activity indicator in distant stars.

Based on the parallax measurement from \emph{Gaia}, \thisstar\ is $230.41 \pm 28.14$ pc away. 
Our line of sight to \thisstar\ crosses three ISM clouds, labeled
``LIC" (-11.49 $\pm$ 1.29 km.s$^{-1}$), 
``G"   (-13.63 $\pm$ 0.97 km.s$^{-1}$), and 
``Mic" (-19.15 $\pm$ 1.38 km.s$^{-1}$)
in \cite{2008ApJ...673..283R}. 
These range from roughly 20 to 30 km.s$^{-1}$ redward of \thisstar's barycentric velocity of -40 km.s$^{-1}$, which may lead to significant ISM absorption if the Ca {\sc ii} column density in the ISM clouds ($\log n_{\rm Ca {\sc II}}$) is high.
Using the calibrations of H{\sc i} column density from E(B-V) of \cite{1994ApJS...93..211D}  and the Ca{\sc ii}/H{\sc i} column density ratio calibration of  \cite{2000ApJ...544L.107W}, we deduced a column density $\log n_{\rm Ca II}$ = 12 $\pm$ 1. According to \cite{Fossati2017}, this is on the edge of being significant.
We visually inspected the Ca {\sc ii} lines, as well as the Na D region (which often shows interstellar absorption) in our HARPS-N spectra of \thisstar, using the spectrum display facilities of the Data \& Analysis Center for Exoplanets\footnote{\url{https://dace.unige.ch}}. We see two  absorption features in the Na D1 and D2 lines at velocities consistent with those of the G and Mic clouds. The stronger of the two features is likely to be associated with the G cloud, which is the furthest away from the barycentric velocity of \thisstar. There are no visible ISM features closer to the stellar velocity, so we conclude that we should not expect the $\log R'_{\rm HK}$ index to be affected significantly by ISM absorption.


\subsubsection{Photometric rotational modulation}\label{photom_mod}
As can be seen in Figure~\ref{fig:ACFplot}, the $Kepler$ light curve is generally quiet but does present occasional bursts of activity, lasting for a few stellar rotations (determined in Section~\ref{arlifetime}). These photometric variations are likely to be the signature of a group of starspots emerging on the stellar photosphere. On the Sun, dark spots by themselves do not induce very large RV variations (of order 0.1-1 m.s$^{-1}$; see \cite{Lagrange2010I,2016MNRAS.457.3637H}). However, they are normally associated with facular regions, which induce significant RV variations \emph{via} the suppression of convective blueshift (on order of the m.s$^{-1}$; see \cite{2010A&A...512A..39M,2010A&A...519A..66M,2016MNRAS.457.3637H}). Therefore, we might still expect to see some activity-driven RV variations over the span of our RV observations, which could eventually affect the reliability of our mass determination for \thisplanet.

\subsection{Determining the rotation period $P_{\rm rot}$ and active-region lifetime $\tau_{\rm ev}$ of the host star}
\label{arlifetime}

We estimated the rotation period and the average lifetime of the starspots present on the stellar surface by performing an autocorrelation-based analysis on the out-of-transit PDCSAP light curve.
We produced the autocorrelation function (ACF) by introducing discrete time lags, as described by \cite{Edelson1988}, in the light curve and cross-correlating the shifted light curves with the original, unshifted curve. 
The ACF resembles an underdamped, simple harmonic oscillator, which we fit via an MCMC procedure. We refer the reader to \cite{2017MNRAS.472.1618G} for further detail on this technique.

The full light curve is shown in the top panel of Figure~\ref{fig:ACFplot}.  As discussed in Section~\ref{photom_mod}, \thisstar\ is relatively quiet and most of the light curve displays no significant rotational modulation. 
We initially computed the ACF of the full out-of-transit PDCSAP light curve, but found it to be flat, thus providing no useful information about the rotation period and active-region lifetime.

We then split the light curve into individual chunks according to their activity levels:
\begin{itemize}
\item{Active light curve:} 
we see occasional ``bursts" of activity, notably in the first 200 days of the lightcurve, which we zoom in on in the middle panel of Figure~\ref{fig:ACFplot}. This photometric variability is visible in both the PDCSAP and PCA light curves (see Figure~\ref{lcs}); the PCA lightcurve has a slightly higher point-to-point scatter likely as a result of a larger aperture. This ``active" chunk spans several \emph{Kepler} quarters, making it unlikely to be the product of quarter-to-quarter systematics. The corresponding ACF is shown alongside it, and our analysis results in a rotation period of 13.8 $\pm$ 0.1 days, and an active-region lifetime of 23 $\pm$ 8 days. 

\item{Quiet light curve:}
the bottom panels of Figure~\ref{fig:ACFplot} show a 400-day stretch of quiet photometric activity, spanning several quarters. The PCA (and DV) lightcurves do not display any variability either. The corresponding ACF analysis yields a rotation period of 12.7$\pm$0.1 days, and an active-region lifetime of 12.2 $\pm$ 2.8 days. 

\end{itemize}

Our rotation period estimates are in rough agreement with each other, although they do differ by more than 1-$\sigma$ according to our MCMC-derived errors. Several factors are likely to be contributing to this. 
First and foremost, the tracers of the stellar rotation, namely the active regions on the photosphere, have finite lifetimes and are therefore imperfect tracers. 
An active region may appear at a given longitude and disappear after a rotation or two, only to be replaced by a different region at a different longitude. These phase changes modulate the period of the activity-induced signal, therefore resulting in a \emph{distribution} of rotation periods as opposed to a clean, well-defined period.
Second, the stellar surface is likely to be dominated by different types of features when it is active and non-active; \emph{eg.} when no spots are present we may be measuring the rotation period induced by bright faculae. In the case of the Sun, it is known that sunspots rotate slightly faster than the surrounding photosphere (see \cite{2004soas.book.....F} and references therein). Following different tracers could plausibly result in differing rotation periods.
Thirdly, we note that differential rotation is often invoked to explain this range in measured rotation periods. While it does have this splitting effect, it is not significantly detectable in \emph{Kepler} light curves of Sun-like stars \citep{2015MNRAS.450.3211A}.

We take the rotation period to be the average value of the estimates we obtained for the various parts of the light curve, and its 1-$\sigma$ uncertainty as the difference between the highest and lowest values we obtained in order to better reflect the range of rotation rates of the stellar surface. This corresponds to a value $P_{\rm rot}$~=~13.6 $\pm$ 1.4 days.

Similarly, the active-region lifetime estimate that we obtain for the quiet light curve is much shorter than that measured in the active portion. At quieter times, the largest spots (or spot groups) will be smaller and will therefore decay  faster than their larger counterparts (see \citet{2017MNRAS.472.1618G}; and \citet{Petrovay:1997fe} among others). 
For the purpose of our RV analysis we choose the longer active-region lifetime estimate of 23 $\pm$ 8 days. 
In Section~\ref{gpar}, we show that varying this value has no significant impact on our planet mass determination.

We note that the rotation period that we measure via this ACF method is in good agreement with the forest of peaks seen in the periodogram of the light curve (see panel (a) of Figure~\ref{fig:peri}).
These photometrically-determined rotation periods fall within the range derived from the $v~\sin~i$ and inclination measurements of \thisstar\, of $(9.2 - 14.8) \pm 2.4$ days (see Section~\ref{analysis}). They are also in agreement with the photometric rotation period determined by \cite{McQuillan2014}, of $15.78 \pm 2.12$ days.

\subsubsection{Sampling of the observations}\label{sampling}

The way the observations are sampled in time can produce ``ghost" signals (\emph{eg.} see \cite{2016MNRAS.456L...6R}). Such spurious signals can significantly impact planet mass determinations, and, in cases where we do not know for certain that the planet exists (\emph{i.e.} we do not have transit observations) they may even result in false detections (as was the case for Alpha Cen B``b"; \cite{2016MNRAS.456L...6R}).
In the paragraphs below, we describe and implement two analytical tools, namely the window function and stacked periodograms. We use them to assess the adequacy of the cadence of the HARPS-N observations and to identify the dominant signals in the dataset.
\

\paragraph{Window function}\label{wf}

\begin{figure*}[t]
\centering
\includegraphics[scale=0.55]{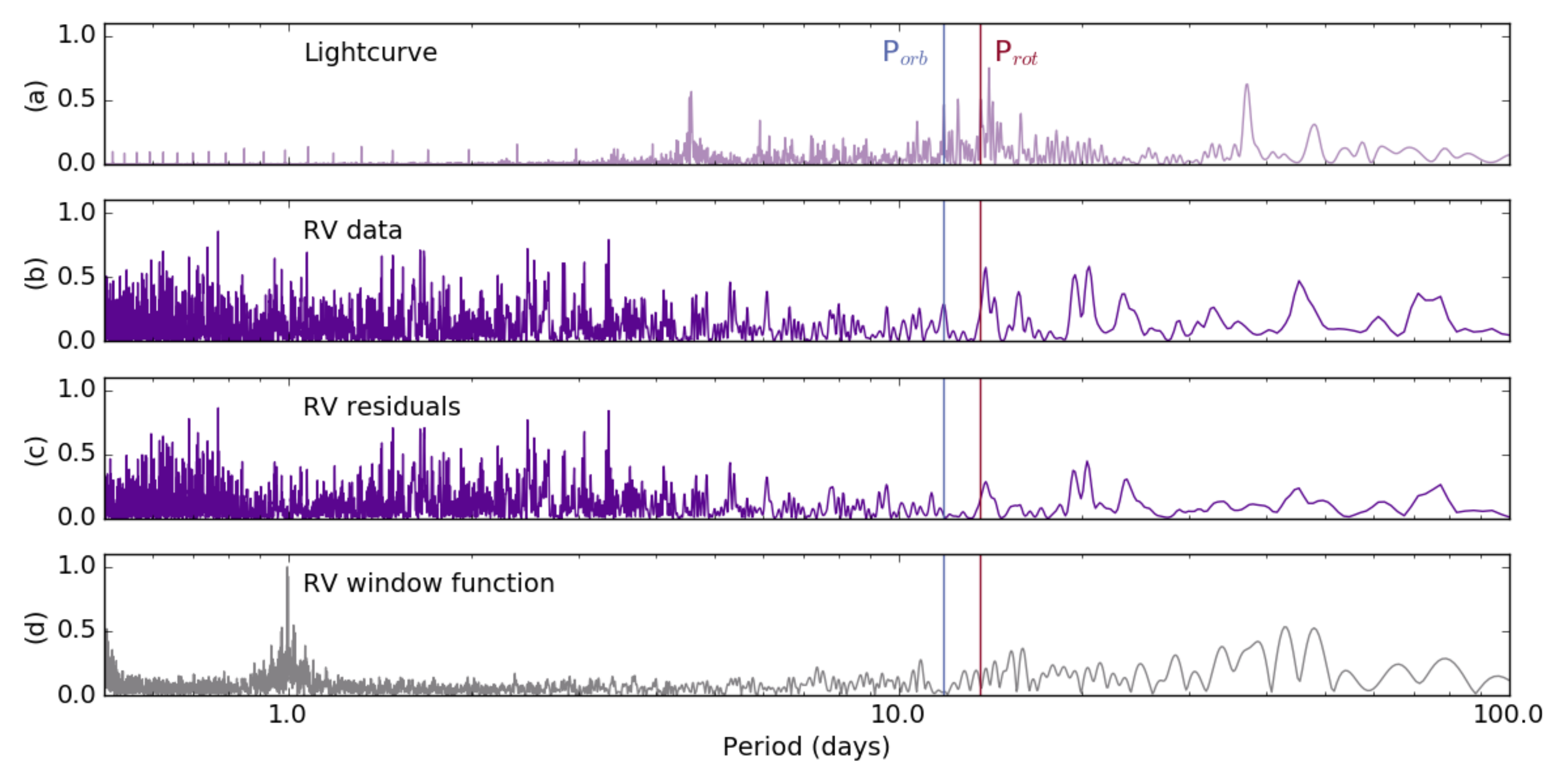}
\caption{Lomb-Scargle periodograms of: (a) the \emph{Kepler} PDCSAP light curve; (b) the HARPS-N RV campaign; (c) the residuals from the RV fit to the HARPS-N observations; and (d) the window function of the RV campaign. None of the peaks in the periodograms of RV observations and residuals are statistically significant (the 99\% false alarm probability levels are higher than the maximum power plotted).
\label{fig:peri}}
\end{figure*}

A simple and qualitatively useful diagnostic is to plot the periodogram of the window function of the observations, as is shown in panel (d) of Figure~\ref{fig:peri}. It is simply the periodogram of a time series with the same time stamps as the RV observations, but with no signals or  noise in the data (\emph{i.e.} the RVs are set to a constant). The observed signal is the convolution of the window function with the real signal. As we might expect, we see a strong forest of peaks centered at 1 day, as a result of the ground-based nature of the observations. 
The highest peak after 1 day is at about 42 days. We note that the 42-day aliases\footnote{See \cite{2010ApJ...722..937D} on calculating aliases.} of the stellar rotation period (of 13.6 days) are 20.1 and 10.3 days. This second alias is rather close to the planet's orbital period, and so we should exercise caution.
This peak around 42 days arises from the fact that past HARPS-N GTO runs have tended to be scheduled in monthly blocks. Regular monthly-scheduled runs can potentially lead to trouble as RV surveys are typically geared towards Sun-like stars, which have rotation periods of about a month; the observational sampling, convolved with the rotationally-modulated activity signals of the star will likely generate beating, spurious signals. 
Fortunately, \thisstar\, has a much shorter rotation period than one month.

\paragraph{Sampling over the rotation period}\label{sampling}
We must also think about whether the time span and cadence of the observations will enable us to sample the stellar rotation cycle densely enough to reconstruct the form of the RV modulation at all phases.

The physical processes and phenomena taking place on the stellar surface undoubtedly result in \emph{signals} with an intrinsic correlation structure (as opposed to random, Gaussian \emph{noise}). 
Typically, they are modulated with the stellar rotation period. The active regions evolve and change over a characteristic timescale (usually a few rotation periods), which changes the phase of the activity-induced signals. If our observations sample the stellar rotation too sparsely, we may not be able to identify these phase-changing, quasi-periodic signals, and recover their real, underlying correlation structure. In this case the signals become noise; their correlation properties may be damped or changed. The sampling may be so sparse that the correlation structure becomes lost completely, in which case the resulting noise will be best accounted for via an uncorrelated, Gaussian noise term (as was the case for Kepler-21 in \cite{LopezMorales2016}).

We obtained 95 observations over 526 nights. This corresponds to 45 orbital cycles and approximately 37 stellar rotation cycles. The two seasons cover about 150 and 200 nights, respectively. This sampling is fairly sparse, and indeed the results of our RV fitting  reflect this (Section~\ref{gpmodels}). 

\begin{figure*}[t]
\centering
\includegraphics[width=\textwidth]{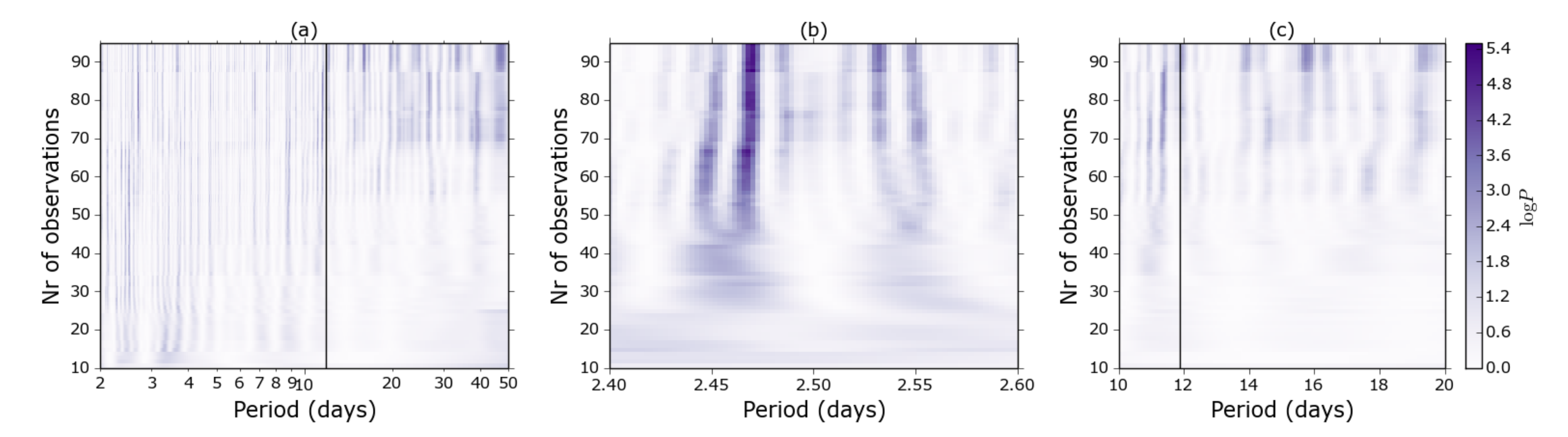}
\centering
\caption{Stacked  periodograms of the HARPS-N RV observations, showing the evolution of the dataset as we gathered the observations. Each panel spans the following period range: (a) 2 to 50 days, (b) zoom-in around 2.5 days, and (c) from 10 to 20 days. The color scale, equal for all three panels represents the periodogram power. Note that the orbital period  (marked by the vertical black line) is at $\perplb$  $\pm$ $ \uperplb $ days and the stellar rotation period is 13.6 $\pm$ 1.4 days. \label{fig:stacked}}
\end{figure*}

\paragraph{Stacked periodograms}
Figure~\ref{fig:stacked} shows the evolution in the Bayesian Generalised Lomb-Scargle periodograms of the RVs as we add more observations \citep{Mortier:2015gw,2017arXiv170203885M}. 
After about 50 observations we begin to see clear power at the orbital period of \thisplanet\ (11.8 days). We note that this is not the only or the most prominent feature in the periodograms.
We also see several streaks of power in the region of 14 to 16 days. This broad range of periods, centered at the rotation period (\prot\ days) is consistent with the relatively short-lived, phase-changing incoherent signatures of magnetic activity. 
We note that these signals are convolved with the window function of the observations, which contains many peaks ranging from about 10 to 50 days (panel (d) of Figure~\ref{fig:peri}).

\paragraph{Periodicities near 2.5 and 3.2 days}
We see strong peaks in the periodograms at periods of 2.5 and 3.2 days. We computed the 99\% and 99.9\% false alarm probability levels via bootstrapping and found that both levels lie well above the highest peaks in the periodograms of both the RV observations and the RV residuals. These signals are therefore not statistically significant. Since we do not have any other information about their nature we did not investigate them any further.

\subsection{Choice of RV model and priors}

\begin{table*}[]
\centering
\caption{Parameters modeled in the RV analysis and their prior probability distributions. \label{tab:priors}}

\begin{tabular}{ccc}
\rule{0pt}{0ex} \\
\hline
\hline
 \rule{0pt}{0ex} \\

Orbital period (from transits) & $P$			&
Gaussian	($11.8728787$, $0.0000085$)			\\

Transit ephemeris (from transits)	& $t_{\rm 0}$			&
Gaussian	($2455013.89795$, $0.00069$)		\\
										
RV semi-amplitude	& $K$					&
Uniform [0, $\infty$]				\\

Orbital eccentricity		& 	 $e$			&
Uniform [0, 1]			\\
		
Argument of periastron	& 	$\omega$	&
Uniform [$0, 2\pi$]							\\
	
Amplitude of covariance & $\eta_1$&
Uniform [0, $\infty$]				\\

Evolution timescale (from ACF) & $\eta_2$&
Gaussian	($23, 8$)		\\

Recurrence timescale (from ACF) & $\eta_3$&
Gaussian	($13.6, 1.4$)		\\

Structure parameter & $\eta_4$&
Gaussian	(0.5, 0.05)		\\

Uncorrelated noise term & $\sigma_{\rm s}$	& 
Uniform [0, $\infty$]								\\	

Systematic RV offset & $RV_{0}$	&
Uniform 									\\

\rule{0pt}{0ex} \\
\hline
\rule{0pt}{0ex} \\

\end{tabular}

\raggedright{For the Gaussian priors, the terms within parentheses represent the mean and standard deviation of the distribution. The terms within square brackets stand for the lower and upper limit of the specified distribution; if no interval is given, no limits are placed.}

\end{table*}


















In light of these preliminary investigations, we choose to stay open to the possible presence of correlated RV noise arising from \thisstar's magnetic activity. We take any such variations into account via Gaussian-process (GP) regression.  Our approach is very similar to that of \cite{LopezMorales2016}. The GP is encoded by a quasi-periodic kernel of the form:

\begin{equation}
\label{cov}
k (t, t') = 
\eta_{1}^{2} \, \cdot \, 
\exp \left[ - \frac{(t - t')^{2}}{2\, \eta_{2}^{2}}  - \frac{2\, \sin^{2} \left( \frac{\pi (t - t')}{ \eta_{3}} \right)}{\eta_{4}^{2}}  \right].
\end{equation}

\noindent The hyperparameter ${\eta_1}$ is the amplitude of the correlated noise; $\eta_2$ corresponds to the evolution timescale of features on the stellar surface that produce activity-induced RV variations; $\eta_3$ is equivalent to the stellar rotation period; and $\eta_4$  gives a measure of the level of high-frequency structure in the GP model.

$\eta_2$ and $\eta_3$ are constrained with Gaussian priors using the values for the stellar rotation period and the active-region lifetime determined via the ACF analysis described in Section~\ref{arlifetime}.

We constrain $\eta_4$ with a Gaussian prior centered around 0.5 $\pm$ 0.05. This value, which is adopted based on experience from previous datasets (including CoRoT-7 \cite{Haywood2014}, Kepler-78 \cite{Grunblatt:2015vi} and Kepler-21 \cite{LopezMorales2016}), allows the RV curve to have up to two or three maxima and minima per rotation, as is typical of stellar light curves and RV curves (see \cite{Jeffers:2009ig}). Foreshortening and limb darkening act to smooth stellar photometric and RV variations, which means that a curve with more than 2-3 peaks per rotation cycle would be unphysical.

\begin{figure*}[t]
\centering
\includegraphics[width=\textwidth]{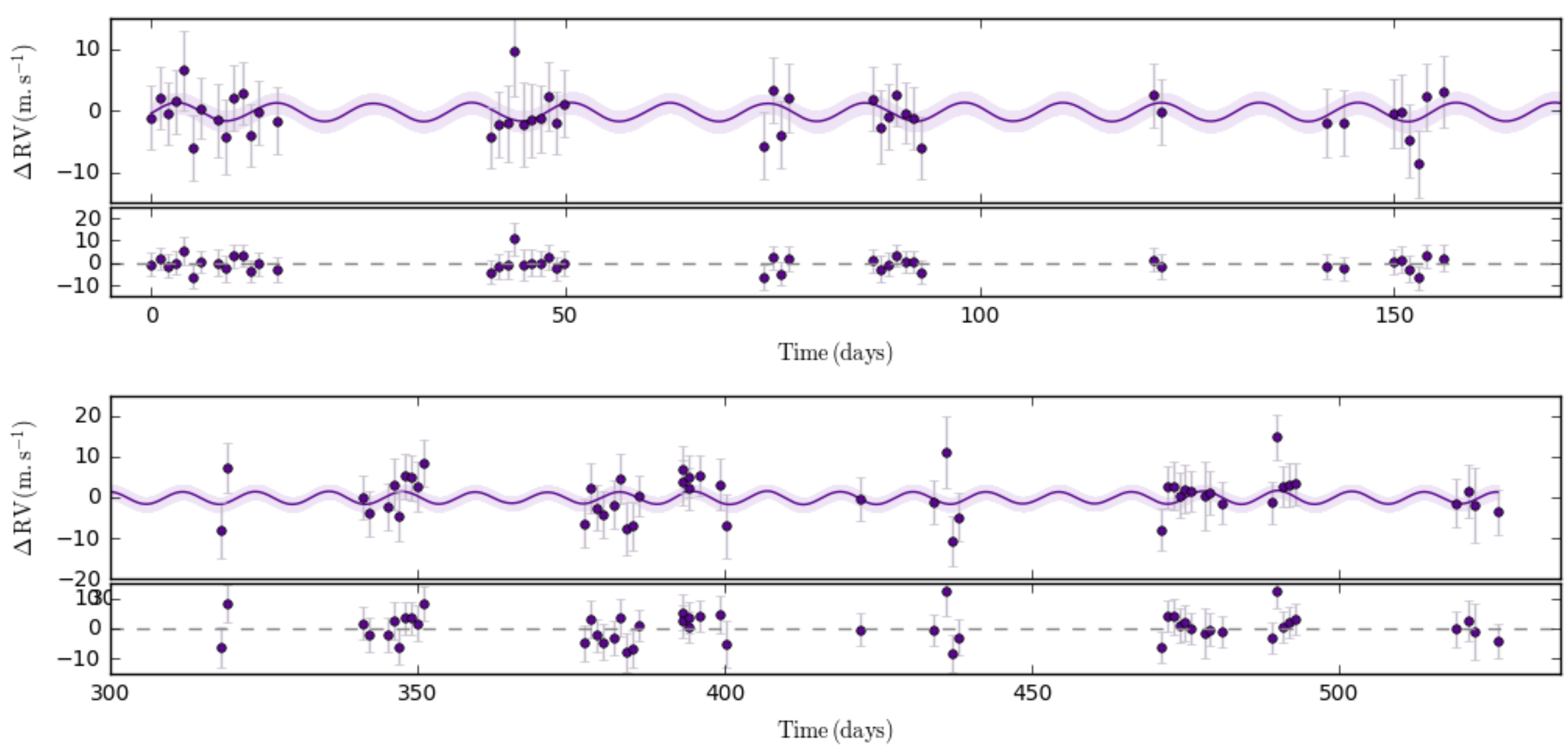}
\centering
\caption{The HARPS-N RV data (points with error bars) and our best fit (dark line with light-shaded 1-$\sigma$ error regions): \emph{(top)} zoom-in on the first season; \emph{(bottom)} zoom-in on the second season, the following year. The residuals after subtracting the model from the data (in m.s$^{-1}$) are shown in the plots below each fit. \label{all}}
\end{figure*}

The strong constraints on the hyperparameters (particularly $\eta_4$) are ultimately incorporated into the likelihood of our model, and as shown in Figures~\ref{all} and~\ref{phase} provide a realistic fit to the activity-induced variations.
We note that GP regression, despite being robust is also extremely flexible. Our aim is not to test how well an unconstrained GP can fit the data, but rather to constrain it to the maximum of our prior knowledge, in order to account for activity-driven signals as best as we can.

We account for the potential presence of uncorrelated, Gaussian noise by adding a term $\sigma_{\rm s}$ in quadrature to the RV error bars provided by the DRS.

We model the orbit of \thisplanet\ as a Keplerian with free eccentricity. We adopt Gaussian priors for the orbital period and transit phase, using the best-fit values for these parameters estimated in Section~\ref{subsectrans}. Finally, we account for the star's systemic velocity and the instrumental zero-point offset of the HARPS-N spectrograph with a constant term $RV_0$. We summarize the priors used for each free parameter of our RV model in Table~\ref{tab:priors}.

The covariance kernel of Equation~\ref{cov} is used to construct the covariance matrix {\bf K}, of size $n$ x $n$ where $n$ is the number of RV observations. Each element of the covariance matrix tells us about how much each pair of RV data are correlated with each other.

For a dataset $\underline{y}$ (with $n$ elements $y_i$), the likelihood $\mathcal{L}$ is calculated as \citep{Rasmussen:2006vz}:
\begin{equation}\label{logl}
\begin {aligned}
\log \mathcal{L} = - \frac{n}{2} \log(2\pi) - \frac{1}{2} \log (\vert {\bf K} + \sigma_i^2\, {\bf I}\vert) \\
- \frac{1}{2} \, \underline{y}^{T}.\, ({\bf K}+ \sigma_i^2 \,{\bf I})^{-1}. \,\underline{y}.
\\
\end{aligned}
\end{equation}
The first term is a normalisation constant. The second term, where $\vert {\bf K} \vert$ is the determinant of the covariance matrix, acts to penalise complex models. The third term represents the $\chi^2$ of the fit. 
The white noise component, $\sigma_i$, includes the intrinsic variance of each observation (\emph{i.e.} the error bar, see Table~\ref{specdata}) and the uncorrelated Gaussian noise term $\sigma_s$ mentioned previously, added together in quadrature. ${\bf I}$ is an identity matrix of size $n$ x $n$.

We maximize the likelihood of our model and determine the best-fit parameter values through a MCMC procedure similar to the one described in \cite{Haywood2014}, in an affine-invariant framework \citep{goodman}.

\subsection{Underlying assumptions in our choice of covariance kernel}
In imposing strong priors on ${\eta_2}$ and ${\eta_3}$, we are making the assumption that the rotation period $P_{\rm rot}$ and active-region lifetime $\tau_{\rm ev}$ are the same in both the photometric light curve and the RV curve. This is potentially not the case, as the photometric and spectroscopic variations may be driven by different stellar surface markers/phenomena (\emph{eg.} starspots, faculae). They may rotate at different speeds, be located at significantly different latitudes on the stellar surface or have very different lifetimes. Faculae on the Sun persist longer than spots. They are likely the dominant contributors to the RV signal, while the shorter-lived spots will dominate the photometry. \\
It is difficult to check the validity of this assumption as these very same factors  also impede our ability to determine precise estimates for $P_{\rm rot}$ and $\tau_{\rm ev}$, particularly in RV observations for which we do not benefit from long-term, high-cadence sampling. For example, the rotation period usually appears in the periodograms (of the light curve and the RVs) as a forest of peaks rather than a single clean, sharp peak (see Figure~\ref{fig:peri}). This effect is the result of the tracers (spots, faculae, \emph{etc.}) having lifetimes of just a few rotations, and subsequently reappearing at different longitudes on the stellar surface. This scrambles the phase and thus modulates the period; see Section~\ref{arlifetime}.

\subsection{Results of the RV fitting}\label{gp}\label{gpmodels}

We investigated the effect of including a GP and/or an uncorrelated noise term on the accuracy and precision of our mass determination for \thisplanet.
We also looked at the effects of using different priors for $\eta_2$ and $\eta_3$, and injecting a fake planet with the density of Earth.

\begin{table*}[]
\centering
\caption{Effects of including correlated and/or uncorrelated noise contributions in our RV fitting. \label{rvtests}}

\begin{tabular}{cccc}
\rule{0pt}{0ex} \\
\hline
\hline
\rule{0pt}{0ex} \\

 & $K$ & $\eta_1$ & $\sigma_{\rm s}$ \\
 & [m.s$^{-1}$] & [m.s$^{-1}$] & [m.s$^{-1}$] \\
 \hline

\rule{0pt}{0ex} \\
\multicolumn{4}{l}{\emph{(a) Original RV dataset}} \\
\rule{0pt}{0ex} \\
Model 1 & 1.47 $\pm ^{ 0.88} _{ 0.80  }$     
& 1.6 $\pm ^{ 1.3} _{ 1.0  }$ 
& 4.3  $\pm$ 0.8  \\

Model 2 & 1.46 $\pm ^{0.81 } _{ 0.74 }$      
&   -            
& 4.6 $\pm$ 0.9     \\

Model 3 & 1.51 $\pm ^{ 0.47  } _{  0.46   }$  
&     -  
&    -           \\

    & & \\

\multicolumn{4}{l}{\emph{(b) RV dataset with injected Earth-composition \thisplanet}} \\
\rule{0pt}{0ex} \\
Model 1 & 6.13 $\pm ^{ 0.90 } _{  0.93  }$    
& 1.8 $\pm ^{1.3  } _{ 1.1 }$    
& 4.4   $\pm ^{0.9 } _{ 0.8 }$     \\

Model 2 & 6.19 $\pm ^{ 0.83  } _{  0.84   }$     
&     -       
& 4.6  $\pm ^{ 0.9  } _{  0.7   }$    \\

Model 3 & 6.25 $\pm$0.48    
&      -
& -\\

    & & \\

\rule{0pt}{0ex} \\
\hline
\rule{0pt}{0ex} \\

\end{tabular}

\raggedright{Model 1: correlated \& uncorrelated noise (GP, $\sigma_{\rm s}$, $RV_0$ and a Keplerian orbit);
Model 2: uncorrelated noise ($\sigma_{\rm s}$, $RV_0$ and a Keplerian orbit);
Model 3: no noise components ($RV_0$ and a Keplerian orbit).
(b): we injected a Keplerian signal with semi-amplitude 6.2 m.s$^{-1}$ (after subtracting the detected amplitude of 1.47 m.s$^{-1}$), corresponding to a mass of 22.6 \mearth. }

\end{table*}

We tested three models accounting for both correlated and uncorrelated noise. The first one, which we refer to as Model 1, contains both correlated and uncorrelated noise, in the form of a GP and a term $\sigma_{\rm s}$ added in quadrature to the errors bars, respectively. In addition, the model has a term $RV_0$ and a Keplerian orbit.
The second model we tested (Model 2) has no GP but does account for uncorrelated noise via a term $\sigma_{\rm s}$ added in quadrature to the error bars. Again, the model also has a term $RV_0$ and a Keplerian orbit.
Our third and simplest model (Model 3) contains no noise components at all. It only contains a zero offset $RV_0$ and a Keplerian orbit.

For all models, we used the same prior values and 1-$\sigma$ uncertainties for all the timescale parameters (orbital $P$ and $t_0$, stellar $P_{\rm rot}$ and $\tau_{\rm ev}$) as well as the structure hyperparameter $\eta_4$. We found that the eccentricity and argument of periastron remained the same in all cases (consistent with a circular orbit). The zero offset $RV_0$ was also unaffected. 
The best-fit values for the remaining parameters ($K$, $\eta_1$, $\sigma_{\rm s}$) for each model tested are reported in Table~\ref{rvtests}.

\begin{table*}[]
\centering
\caption{System Parameters for the \thisstar\ system. \label{bigtable}\label{tab:star}}

\begin{tabular}{ccc}
\rule{0pt}{0ex} \\
\hline
\hline
\rule{0pt}{0ex} \\

Parameter & Value & Source \\
\rule{0pt}{0ex} \\
\hline

	RA [h m s]      & 19 06 45.44 &  a \\
	DEC [d m s]     & +39 12 42.63 & a \\
	Spectral type   & G0V  \\
	$m_V$   & 11.05$\pm$0.08 &a \\
	$B-V$   & 0.57 & a\\
	Parallax [mas]  & 4.34$\pm$  0.53 & b \\
	Distance [pc]   & 230.41$\pm$ 28.14 & \\
	$T_{\rm eff}$ [K]  & 6148$\pm$ 71 & c \\
	$\log{g}$       & 4.36$\pm$ 0.10 & c \\
	${\rm [Fe/H]}$  &$-$0.24$\pm$ 0.05 & c \\
	$\Delta\nu$ [$\mu$Hz] & 128.8$\pm$ 1.3 & d \\
	$\nu_{max}$ [$\mu$Hz] & 2928.0$\pm$ 97.0 & d \\
	Mass $[M_{\odot}]$ & 1.03$\pm$ 0.04 & e \\
	Radius $[R_{\odot}]$ & 1.03$\pm$ 0.02 & e \\
	$\rho_{\ast}$ [$\rho_{\odot}$] & 0.94$\pm$ 0.04 & \\
    Age [Gyr] & 2.56$\pm$1.06 & e \\
	$v\,\sin{i}$ [km~s$^{-1}$]      & 3.5$\pm$ 0.5 & d \\

    Limb darkening $q_1$~ & $\ldone$ $\pm$ $ \uldone$ & f \\
    Limb darkening $q_2$~ & $\ldtwo$ $\pm$ $ \uldtwo$ & f \\	

	$<\log{R'_{\rm HK}}>$ & $-$4.97 & g \\

	P$_{\rm rot}$ [days] & $\prot$  $\pm$  $\prote$ & f \\
	$\tau_{\rm ev}$ [days] & $\tev$ $\pm$  $\teve$ & f \\

\hline

\emph{Transit and radial-velocity parameters} & & \\
    Orbital period $P$~[days] & $\perplb$  $\pm$ $ \uperplb $ & f \\
    Time of mid-transit $t_{0}$~[BJD] & $\ttransitb$  $\pm$ \uttransitb & f\\ 
    Radius ratio $(R_b/R_\star)$ & $\rprstb$   $\pm$ $ \urprstb$ & f \\
    Orbital inclination $i$~[deg] & $\inclb$   $\pm$ $ \uinclb$ & f \\ 
    Transit impact parameter $b$ & $\impb$   $\pm$ $ \uimpb$ & f \\
    RV semi-amplitude $K$ [m.s$^{-1}$] & $\kamp$   $\pm$ $ ^{\kampeu} _{-\kampel}$ & g \\
    RV semi-amplitude 68\% (95\%) upper limit [m.s$^{-1}$] & $<1.8$ ($<2.8$) & g \\
    Eccentricity 68\% (95\%) upper limit &$<0.36$ ($<0.79$)& g \\  
    Argument of periastron $\omega_p$ [deg] & -71 $\pm$ 92 & g \\
    RV offset RV$_0$ [km.s$^{-1}$]   & -40.6386 $\pm$ 0.000006 & g \\
    
                              
\hline

\emph{Derived parameters for \thisplanet} & & \\
    Radius $R_b$~[\rearth] & $\rplb$     $\pm$ $ \urplb$  & e,f \\
    Mass $M_b$~[\mearth] & $\mass$     $\pm$ $ ^{\masseu} _{-\massel}$  & e,f,g \\
    Mass 68\% (95\%) upper limit [\mearth] & $< 6.2$ ($< \mupp$) &        e,f,g \\
    Density $\rho_b$~[g.cm$^{-1}$] & $\rhopl$    $\pm$  $^{\rhopleu} _{-\rhoplel}$  & e,f,g \\
    Density 68\% (95\%) upper limit [g.cm$^{-1}$] & $< 3.2$  ($< 5.1$)  & e,f,g \\
    Scaled semi-major axis $a/R_\star$  & $\arstb$   $\pm$ $\uarstb$ & f \\
    Semi-major axis $a_b$ [AU] & 0.103 $\pm$ 0.001 & f, g \\
    Incident flux $F$~[\fearth] & $\flux$   $\pm$  $\fluxe$  & e,f \\

\rule{0pt}{0ex} \\
\hline
\rule{0pt}{0ex} \\

\end{tabular}

	\tablenotetext{1}{\cite{Hog2000}}
	\tablenotetext{2}{\cite{Gaia2016}}
	\tablenotetext{3}{from ARES+MOOG analysis, with the surface gravity corrected following \cite{Mortier2014}}
	\tablenotetext{4}{\cite{Huber2013}}
	\tablenotetext{5}{Bayesian estimation using the PARSEC isochrones and asteroseismology}
	\tablenotetext{6}{Analysis of the $Kepler$ lightcurve}
	\tablenotetext{7}{Analysis of the HARPS-N spectra/RVs}

\end{table*}

Overall, the value of the RV semi-amplitude of \thisplanet\ is robust to within 5 cm.s$^{-1}$, regardless of whether we account for (un)correlated noise or not. This is a reflection of the fact that the host star has fairly low levels of activity.
When the GP is included, its amplitude $\eta_1$ is similar to that of $K$. 
However, we note that the uncorrelated noise term is large and dominates both the GP and the planet Keplerian signal. This may be a combination of additional instrumental noise (the star is very faint and our observations are largely dominated by photon noise) and short-term granulation motions. Also, rotationally-modulated activity signals that were sampled too sparsely may also appear to be uncorrelated and be absorbed by this term rather than the GP (as was likely the case in \cite{LopezMorales2016}).

Regardless of its nature, we cannot ignore the presence of uncorrelated noise. Doing so would lead us to underestimating our 1-$\sigma$ uncertainty on $K$ by 40\%.
Finally, we see that when we go from Model 2 (uncorrelated noise only) to Model 1 (correlated and uncorrelated noise), the uncertainty on $K$ increases by about 7 cm.s$^{-1}$. We attribute this slight inflation to the fact that the orbital period of \thisplanet\ is close to the rotation period of its host star. This acts to incorporate this proximity of orbital and stellar timescales in the mass determination of \thisplanet.

\begin{figure}[t]
\centering
\includegraphics[scale=0.35]{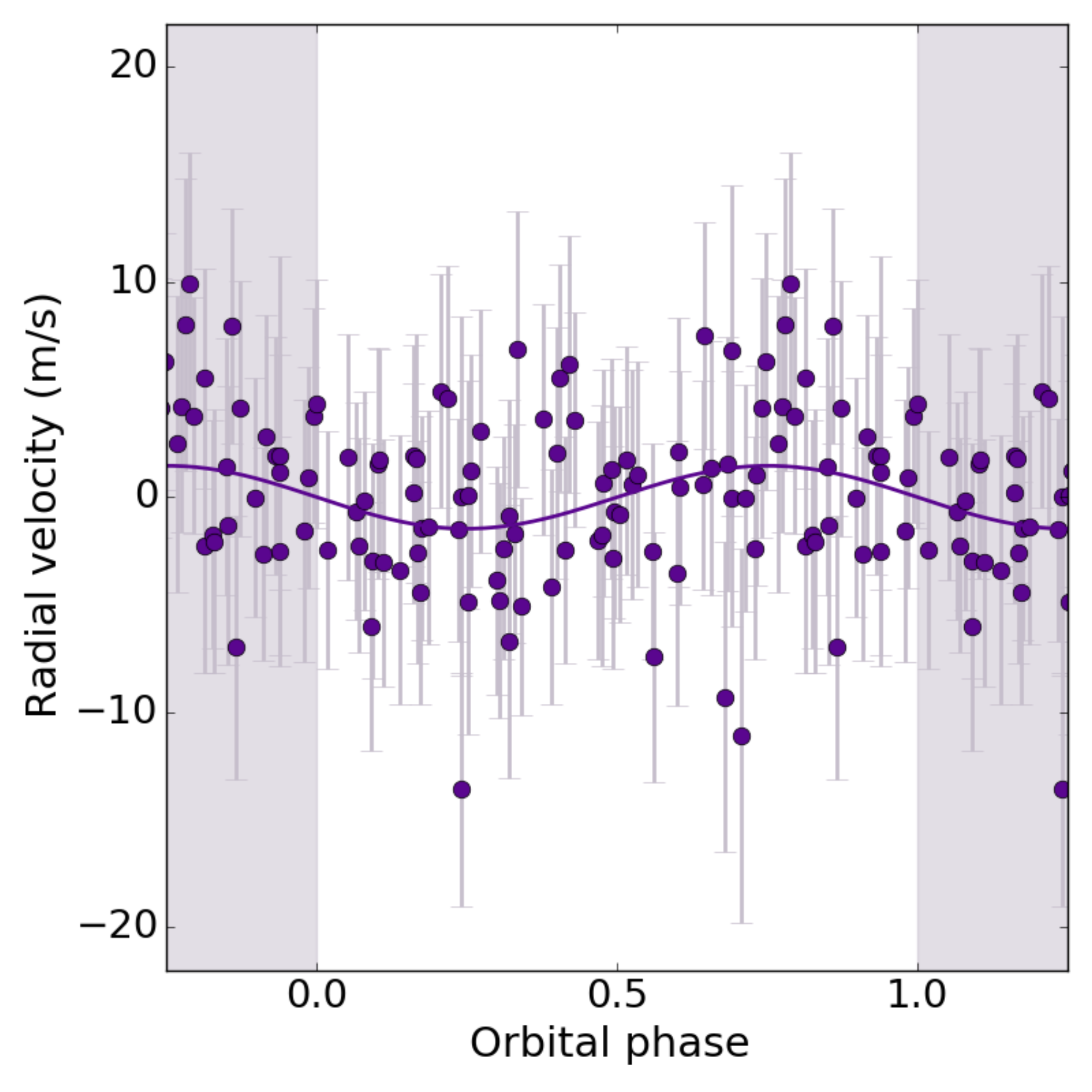}
\caption{Phase plot of the orbit of \thisplanet\ for the best-fit model after subtracting the Gaussian-process component. \label{phase}}
\end{figure}

\subsubsection{Effects of varying $P_{\rm rot}$ and $\tau_{\rm ev}$} \label{gpar}
We ran models with different values for the stellar rotation and evolution timescales ($P_{\rm rot}$ ranging between 11-20 days, $\tau_{\rm ev}$ ranging between 13-50 days, with associated uncertainties ranging from 1-8 days for both). We found that the amplitude of the GP and its associated uncertainty remained the same throughout our simulations. The semi-amplitude of \thisplanet\ also remained the same to within 10\%, ranging between 1.37-1.47 m.s$^{-1}$, with a 1-$\sigma$ uncertainty ranging from 0.85-0.90 m.s$^{-1}$. The uncertainty was largest in cases with the longest evolution timescale (\emph{i.e.}, the activity signals are assumed to retain coherency for a long time) and when the rotation period overlapped most with the orbital period of \thisplanet\ (at 11 days).

\subsubsection{RV signature of \thisplanet\ if it had an Earth-like composition}\label{ifrocky}
We subtracted a Keplerian with a semi-amplitude $K$ of 1.47 m.s$^{-1}$, corresponding to that of Model 1, and subsequently injected a Keplerian signal with semi-amplitude 6.2 m.s$^{-1}$, \emph{i.e.} a mass of 22.6 \mearth (at the period and phase of \thisplanet). With \thisplanet's radius of 2.213 \rearth and according to the composition models of \cite{Zeng2016}, these mass and radius values correspond to an Earth-like composition.
We tested all three Models after injecting this artificial signal.
As shown in panel (b) of Table~\ref{rvtests}, we see a completely consistent behavior when the semi-amplitude of the planet is artificially boosted. In particular, the amplitude $\eta_1$ of the GP remains consistent well within 1-$\sigma$. This artificial signal is detected at high significance (7-$\sigma$).
This test confirms that if the planet had an Earth-like composition, our RV observations would have been sufficient to determine its mass with accuracy and precision; it therefore shows that \thisplanet\ must contain a significant fraction of volatiles.
We find that only 0.014\% of the samples in our actual posterior mass distribution lie at or above 22.6 \mearth, and therefore conclude that we can significantly rule out an Earth-like composition for this planet. 

\begin{figure}[t]
\centering
\includegraphics[scale=0.28]{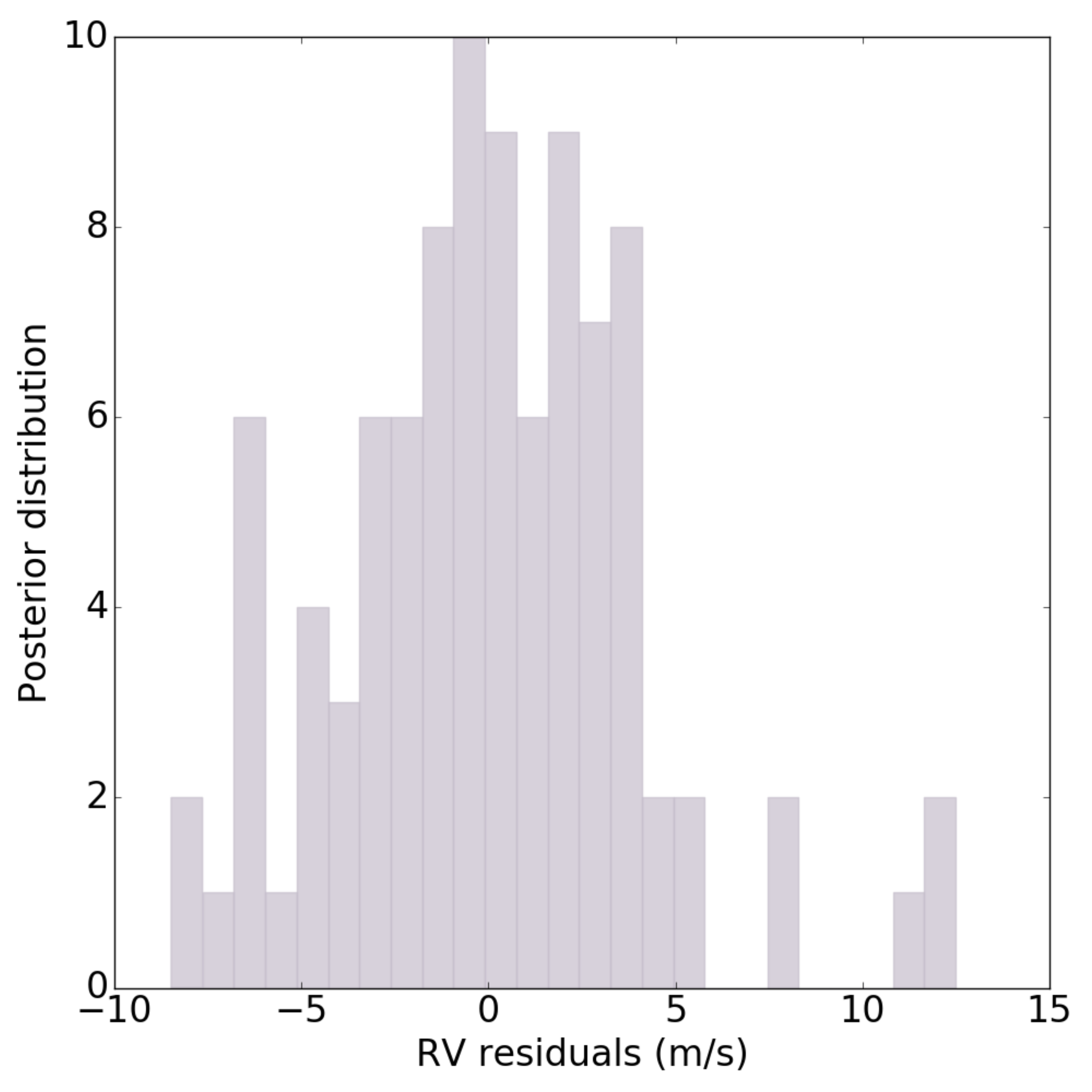}
\caption{Histogram of the residuals of the RVs, after subtracting Model 1 from the data. The residuals are close to Gaussian-distributed. \label{rvresid_histo}}
\end{figure}

\subsection{Mass and composition of \thisplanet}
The RV fit from Model 1, which we adopt for our mass determination, is plotted in Figure~\ref{all}. The corresponding correlation plots for the parameters in the MCMC run, attesting of its efficient exploration and good convergence, are shown in the Appendix. The residuals, shown as a histogram in Figure~\ref{rvresid_histo} are Gaussian-distributed. The phase-folded orbit of \thisplanet\ is shown in Figure~\ref{phase}.

\begin{figure}[t]

\centering
\includegraphics[scale=0.28]{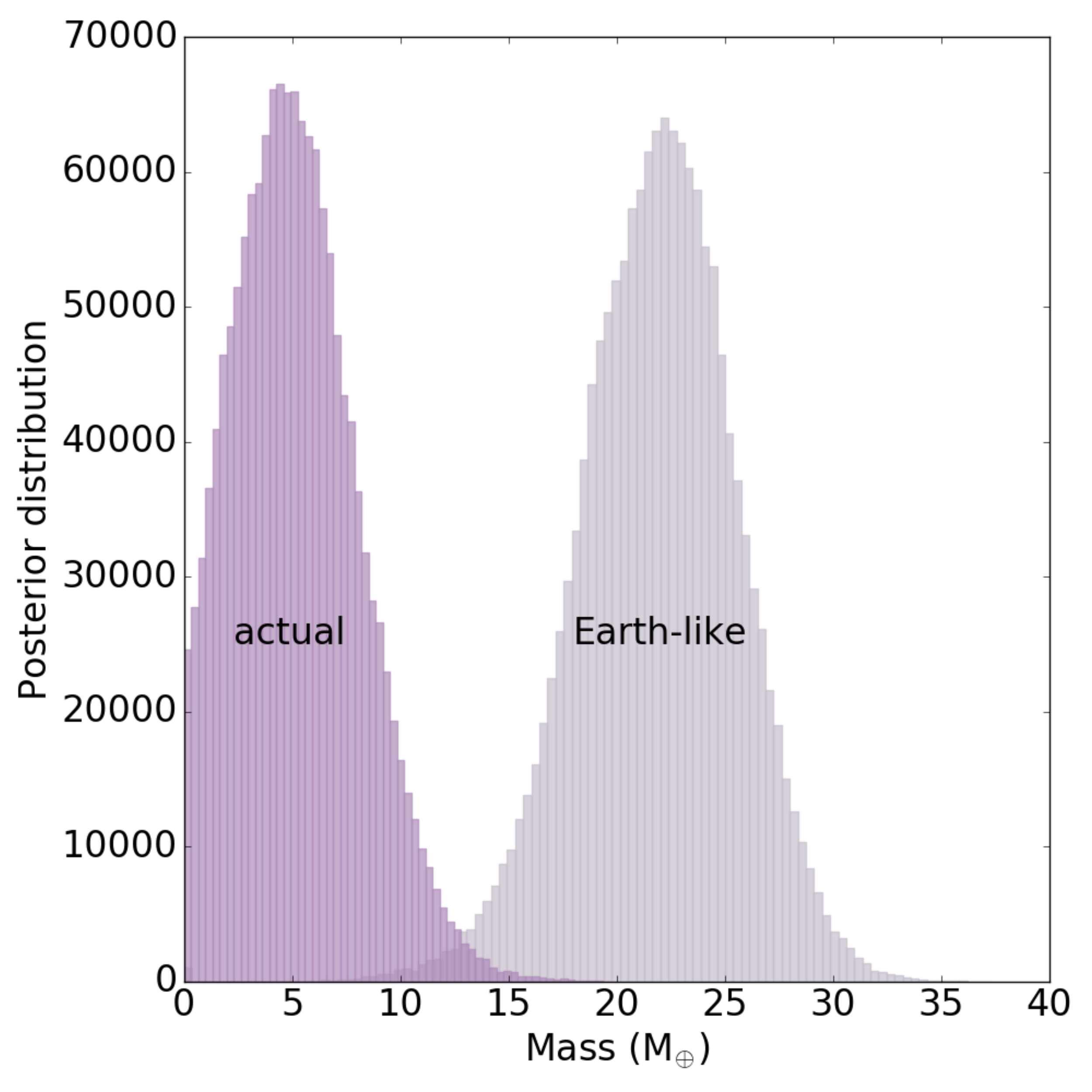}
\caption{Posterior distributions of the mass parameter: ``actual" refers to the  posterior distribution obtained when fitting Model 1 to the actual RV data, while ``Earth-like" is the distribution we obtain if we inject a planet of Earth-like composition with \thisplanet's 2.2 \rearth\ radius. Such a planet would have a mass of $\approx$ 22.6\mearth. \label{mass_histo}}
\end{figure}

Taking the semi-amplitude obtained from Model 1, we determine the mass of \thisplanet\ to be \mass $\pm ^{\masseu} _{\massel}$ \mearth. The posterior distribution of the mass is shown in Figure~\ref{mass_histo}. For comparison, we also show the posterior distribution obtained after we injected the artificial signal corresponding to a \thisplanet \, with an Earth-like composition. As we discussed in Section~\ref{ifrocky}, we see that the two posterior distributions are clearly distinct and with little overlap. Despite the low significance of our planet mass determination, we can state with high confidence that \thisplanet \, has a significant gaseous envelope and is not Earth-like in composition. The mass of \thisplanet \, is less than 6.2 \, \mearth \, at 68\% confidence, and less than \mupp \, \mearth\, at 95\% confidence. Our analysis excludes an Earth-like composition with more than 98\% confidence (see Section~\ref{ifrocky}).

We obtain a bulk density for \thisplanet \ of $\rho_b = 2.5 \pm ^{1.6}_{1.4}$ g.cm$^{-3}$. The planet's density is less than 3.2 g.cm$^{-3}$ to 68\% confidence and less than 5.1 g.cm$^{-3}$ to 95\% confidence.

 The planet may have experienced some moderate levels of evaporation, which may be significant if its mass is indeed below 5 \mearth. 

The eccentricity is consistent with a circular orbit and with the constraints derived from our asterodensity profiling analysis (Section~\ref{asterodensity}). At an orbital period of 11.8 days, we do not expect this planet to be tidally circularized.

The large uncertainty on our mass determination is not unexpected. First, the host star is fainter than typical HARPS-N targets ($m_V = 11.05$) so our RV observations are photon-limited. Second, the window function of the RV observations contains a number of features in the 10-50 day range (see panel (d) of Figure~\ref{fig:peri} and Section~\ref{wf}), which implies that the stellar rotation period, close to 14 days, is sampled rather sparsely. Any activity-induced RV variations, which can reasonably be expected at the level of 1-2 m.s$^{-1}$ from suppression of convective blueshift in facular areas, will thus be sparsely sampled; this is likely to wash out their correlated nature and will result in additional uncorrelated noise -- which in turn inflates the uncertainty of our mass determination.

\section{Discussion: \thisplanet\ among other known exoplanets}\label{discussion}\label{montecarlo}

\begin{figure}[t]
\centering
\includegraphics[scale=0.56]{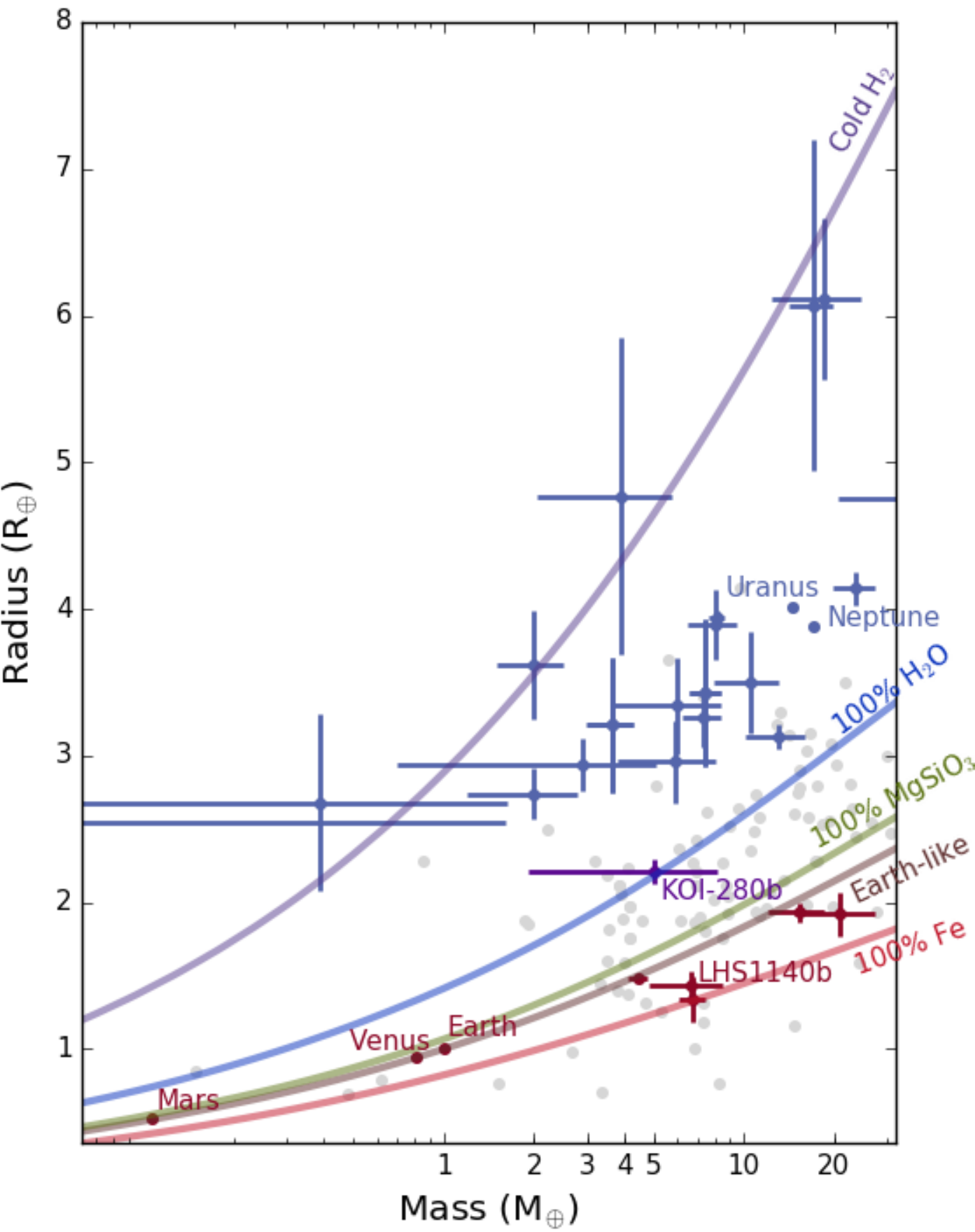}
\caption{Mass-radius diagram for planets in the 0.1--32 \mearth and 0.3--8 \rearth ranges. The blue points correspond to ``gas-rich" planets, while the red points represent planets that are very likely to be rocky in composition (see Section~\ref{montecarlo}). The planets that fall in neither category are colored in gray, and their error bars are omitted for clarity. \thisplanet~ and its associated 1-$\sigma$ measurement uncertainties are shown in purple. \label{massradius}}
\end{figure}

With a radius of 2.213 \rearth \, and a mass less than 10.1 \mearth \, (at 95\% confidence), \thisplanet \, straddles the region between small, rocky worlds and larger, gas-rich worlds.
Figure~\ref{massradius} shows the place of \thisplanet\ as a function of mass and radius, alongside other well-characterized exoplanets in the 0.1--32 \mearth \, and 0.3--8 \rearth \, range.
The exoplanets that are shown have measured masses and were taken from the list compiled by \cite{2017arXiv170601892C}. We used radius measurements from \citet{2017arXiv170310375F} where available, or extracted them from the NASA Exoplanet Archive\footnote{\url{https://exoplanetarchive.ipac.caltech.edu}, operated by the California Institute of Technology, under contract with NASA under the Exoplanet Exploration Program.} otherwise. We include the planets of the solar system, with data from the NASA Goddard Space Flight Center archive\footnote{\url{https://nssdc.gsfc.nasa.gov/planetary/factsheet/}, authored and curated by D. R. Williams at the NASA Goddard Space Flight Center.}.
We overplot the planet composition models of \cite{Zeng2016}.

For the purpose of the present discussion, we identify and highlight the planets that have a strong likelihood of being gaseous (in blue) and rocky (in red). 
For each planet, we drew 1000 random samples from a Gaussian distribution centered at the planet mass and radius, with a width given by their associated mass and radius 1-$\sigma$ uncertainties. 
Planets whose mass and radius determinations indicate a 96\% or higher probability of lying above the 100\% H$_2$O line are colored in blue. Planets that lie below the 100\% MgSiO$_3$ line with 96\% probability or higher, and have a probability of less than 4\% of lying above the 100\% H$_2$O line  are colored in red.
All other planets, colored in gray, are those that do not lie on either extreme of this probability distribution (even though their mass and radius measurement uncertainties may be smaller than others that we identified as rocky or gaseous). For clarity  we omitted their error bars on this plot.

We note that \thisplanet, shown in purple is one of these intermediate worlds.

For this discussion, we define ``water worlds" as planets for which the majority of their content (75-80\% in terms of their radius) is not hydrogen. Their densities indicate that they must have a significant non-rocky component, but this component is water rather than hydrogen. They formed from solids with high mean molecular weight. 
We refer to planets with a radius fraction of hydrogen to core that is greater than 20\% as ``gaseous worlds".

\begin{figure*}[t]
\centering
\includegraphics[scale=0.5]{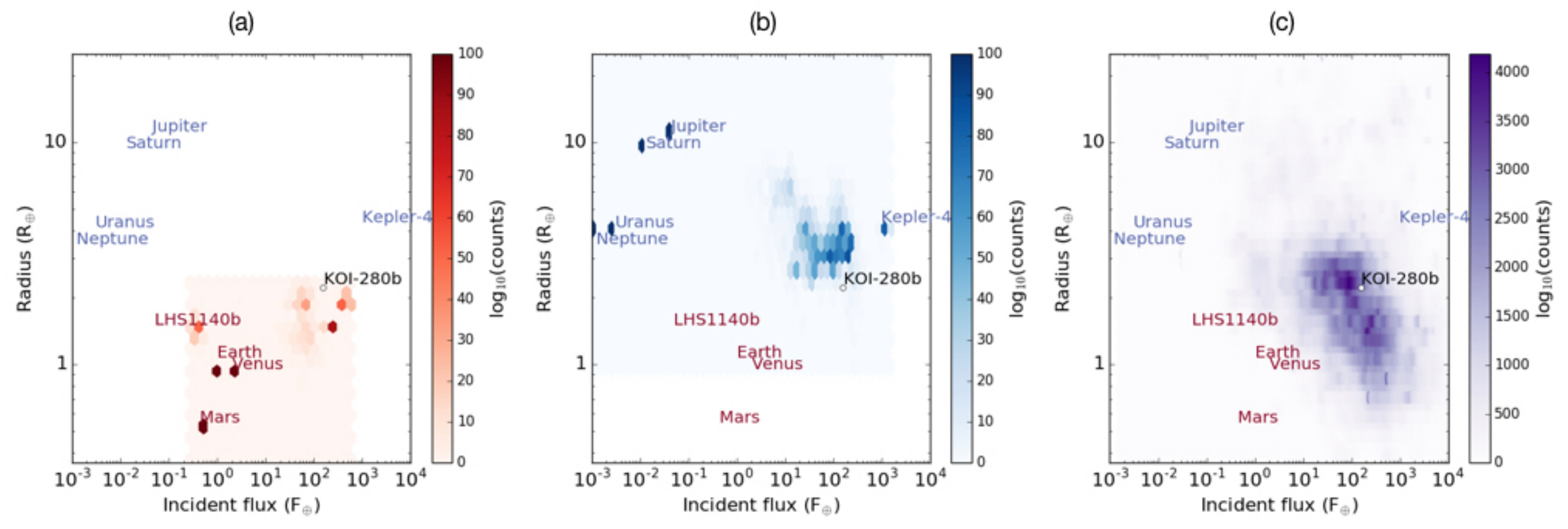}
\caption{Incident flux as a function of radius for well-characterized planets to date.
In all three panels, each planet is represented as a probability-distribution ``cloud" (see main text for details). Panel (a): density distribution for planets that are likely rocky, i.e. with $>$ 96\% probability of lying below the 100\% MgSiO$_3$ line, and $<$ 4\% probability of lying above the 100\% H$_2$O line. Panel (b): density distribution for planets identified as gas-rich, i.e. with $>$ 96 \% probability of lying above the 100\% H$_2$O line. Panel (c): all planets from the  CKS sample \citep{2017arXiv170310375F}, including those without a mass determination. Both the evaporation valley and evaporation desert are apparent. \label{proba_radius}}
\end{figure*}

We wish to investigate how gaseous planets (lying above the water line) behave as a function of planet radius and incident flux received at the planet surface as compared to their rocky counterparts. 
For this purpose, we created the three plots shown in Figure~\ref{proba_radius}, in which planets are again displayed as probability density distributions rather than single points with 1-$\sigma$ uncertainties. For each planet, we draw 1000 random samples from a Gaussian distribution centered at the planet radius and incident flux measurements, with a width given by their associated 1-$\sigma$ uncertainties. We display the resulting distributions as a two-dimensional binned density plot so that the regions of higher probability appear darker. 

In panel (a), we show the resulting density distribution for planets that we previously identified as rocky worlds -- over 96\% of the Gaussian draws fall below the MgSiO$_3$ line and fewer than 4\% fall above the the H$_2$O line.
In panel (b), we show the density distribution for planets that we previously identified as gaseous worlds -- over 96\% of the Gaussian draws fall above the H$_2$O line. 
In both panels (a) and (b), we plot the well-characterized sample described earlier in the discussion (planets with mass determinations listed in \cite{2017arXiv170601892C}, and radius and incident flux measurements from \citet{2017arXiv170310375F} or the NASA Exoplanet Archive). We include the planets of the solar system. We label the planets of the solar system and the planets responsible for some of the more prominent features, as well as the position of \thisplanet~ to guide the reader. 
In panel (c), we show all 2025 planets with updated radii and incident fluxes from the CKS survey \citep{2017arXiv170310375F}. The labels for the solar system planets, LHS1140b and Kepler-4b are included to facilitate comparison with panels (a) and (b).

As has been noted in previous works including \cite{weissmarcy,angie,JontofHutter2016}, we see a great diversity of masses for small, rocky planets (see Figure~\ref{proba_radius}a). They are also present in a broad range of incident fluxes (from $< 1$ \fearth \ up to 10,000 \fearth). Gas-dominated planets also span a wide range of masses, but seem to occur in a narrower range of incident fluxes (see Figure~\ref{proba_radius}b). 
Both rocky and gaseous planets at longer orbital periods, and thus low incident fluxes (below a few \fearth) are more difficult to detect and characterize; this means that our exoplanet sample is most likely incomplete in this flux range. 
We note that Figure~\ref{proba_radius} is not corrected for any such observational biases.
Planets at very high incident fluxes, however, are easiest to detect as they are in very close orbits. 
We note that the known population of hot Jupiters, at large radius and extremely high incident flux (up to 10,000 \fearth) is not represented in these plots; however, previous studies have shown that at the high end of the radius distribution, the hot Jupiters have so much gas that they keep most of it, even in highly-irradiated orbits.

In panel (c) of Figure~\ref{proba_radius}, we see the evaporation valley between 1.5 and 2 \rearth~ that was recently observed by \cite{2017arXiv170310375F} (see also \cite{2017LPI....48.1576Z}) and predicted theoretically by \cite{Owen2017} and \cite{2017arXiv170600251J}.

At intermediate radii, Figure~\ref{proba_radius}c shows a dearth of planets at the highest incident fluxes with radii $\sim$ 2-4 \rearth. It has been shown to be unlikely to be dominated by observational biases, and is commonly referred to as the evaporation desert or sub-Neptune desert \citep{2014ApJ...787...47S,2016NatCo...711201L}.

Planets in this region either do not exist or they are extremely rare. Perhaps it is a transition region, and they will exist in this region but only for a very short time, making them very difficult to detect.

The evaporation desert leads us to speculate on the composition of Neptune-size planets, and their formation and migration histories.
Models of planet interiors are limited by degeneracies in composition for a given mass and radius, regardless of how precisely these two observables may be determined \citep{2010ApJ...712..974R}. This is especially an issue for planets with sizes in the super-Earth to small Neptune range. 
The very existence of the evaporation desert and the evaporation valley argues against a very water-rich population. Water worlds would survive in close-in, highly-irradiated orbits; they could lose their H/He envelopes through evaporation, but the majority of their steam envelopes would remain, so they would never be stripped down to bare rock  \citep{Lopez2016}. However, further studies need to be carried out to understand exactly how strong these constraints are.

On one hand, the distribution of highly-irradiated, low-mass planets is mainly shaped by \emph{formation} processes, such as whether most planets form before their disks dissipate. 
On the other hand, it may be that they are shaped by \emph{evolution} processes, such as evaporation. 
\cite{LopezRice2016} show that constraining the slope of the rocky/non-rocky transition (the edge of the evaporation valley) can differentiate between these two scenarios. 

\thisplanet \ falls in the midst of this transition region, and is in an orbit where the irradiation levels start to be high enough that it is in a relatively unpopulated zone. It is therefore part of a population of planets that we should actively seek to characterize further.

\section{Conclusions}
\label{summary} 
We confirm the planetary nature of \thisplanet, characterize its host star and determine its radius and mass.

Our main conclusions are:

\begin{itemize}
\item
\thisplanet\, is a moderately-irradiated (F = 155 $\pm$ 7 \fearth), sub-Neptune with a substantial gas envelope. We measure its radius to be \rplb$\pm$\urplb \,\rearth, and determine its mass to be \mass$\pm^{\masseu}_{\massel}$ \,\mearth, or less than 10.1 \mearth \, at 95\% confidence. This places \thisplanet\, in a still relatively unexplored area of parameter space, where it straddles the observed evaporation valley between small, rocky planets and Neptune-size, gaseous worlds \citep{2017arXiv170310375F,Owen2017,2017arXiv170600251J}. In addition, its moderately-irradiated orbit places it close to the observed evaporation desert \citep{2014ApJ...787...47S,2016NatCo...711201L}.

\item
The host star \thisstar\, is a G0V Sun-like star with a rotation period of $13.6 \pm 1.4$ days. The magnetic activity behavior that we observe in both our photometric and spectroscopic time series are similar to those of the Sun in its quieter phase. We see the occasional emergence of active regions with average lifetimes of  $23 \pm 8$ days, as measured from the \emph{Kepler} photometric curve via an autocorrelation analysis. We measure activity-driven radial-velocity variations with an RMS of 0.5 m.s$^{-1}$. This value is consistent with ongoing HARPS-N observations of the Sun as a star, that display an RMS of 1.6 m.s$^{-1}$ even though the Sun is now entering the low phase of its 11-year magnetic activity cycle (see \cite{Dumusque:2015ApJ...814L..21D} and Phillips et al., in prep.). Our findings are also consistent with activity levels of order 1-2 m.s$^{-1}$ seen in the quietest main-sequence, Sun-like stars in spectroscopic surveys (\emph{eg.} \cite{Isaacson:2010ud,Motalebi2015}).

\item
In the \thisstar\, system, the radial-velocity RMS induced by magnetic activity, even though it is a relatively quiet star, is of comparable magnitude to the orbital reflex motion induced by the planet \thisplanet.  We account for activity variations as both correlated and uncorrelated noise to obtain an accurate (though not necessarily precise) planetary mass determination. In agreement with previous studies (\emph{eg.} \cite{LopezMorales2016,Rajpaul2015}), we see that the precision of our mass determination depends crucially on regular and adequate sampling of the stellar rotation timescale. If the activity signals are sampled too sparsely, their correlation structure will be changed or lost, in which case they will be best accounted for through an uncorrelated, Gaussian noise term; this will in turn inflate the uncertainty associated with our mass determination.

\item 
It is difficult to measure rotation periods accurately, as they can be different at different levels of activity, likely because the stellar surface is dominated by different types of active regions (\emph{eg.} faculae, spots). For this reason, extra care must be taken in radial-velocity analyses, particularly in systems such as \thisstar, where the stellar rotation ($13.6 \pm 1.4$ days) and planetary orbital period (\perplb$\pm$\uperplb\ days) are close to each other.

\end{itemize}

In order to robustly constrain our planet formation models and look into the details of all these scenarios and processes, we require mass determinations that are accurate and reliable. It is especially important that we focus our characterization efforts on planets like \thisplanet\, that straddle observational boundaries, such as the evaporation valley between gaseous and rocky planets and the evaporation desert at high irradiation levels. Determining the masses of planets like \thisplanet\ is a necessary step to building a statistical sample that will feed models of planetary formation and evolution.

\acknowledgments

We are grateful to Natalie M. Batalha and Samuel N. Quinn for insightful discussions that have helped shape our discussion on planet population properties. We would also like to thank the anonymous referee for providing constructive feedback on the manuscript.
This work was performed in part under contract with the California Institute of Technology (Caltech)/Jet Propulsion Laboratory (JPL) funded by NASA through the Sagan Fellowship Program executed by the NASA Exoplanet Science Institute (R.D.H., C.D.D.).
Some of this work has been carried out within the framework of the NCCR PlanetS, supported by the Swiss National Science Foundation.
 A.V. is supported by the NSF Graduate Research Fellowship, grant No. DGE 1144152.
 A.C.C. acknowledges support from STFC consolidated grant number ST/M001296/1.
 D.W.L. acknowledges partial support from the Kepler mission under NASA Cooperative Agreement NNX13AB58A with the Smithsonian Astrophysical Observatory. 
 X.D. is grateful to the Society in Science-Branco Weiss Fellowship for its financial support. 
 C.A.W. acknowledges support by STFC grant ST/P000312/1.
 This publication was made possible through the support of a grant from the John Templeton Foundation. The opinions expressed are those of the authors and do not necessarily reflect the views of the John Templeton Foundation. 
This material is based upon work supported by the National Aeronautics and Space Administration under grants No. NNX15AC90G and NNX17AB59G issued through the Exoplanets Research Program. The research leading to these results has received funding from the European Union Seventh Framework Programme (FP7/2007-2013) under grant Agreement No. 313014 (ETAEARTH). 
The HARPS-N project has been funded by the Prodex Program of the Swiss Space Office (SSO), the Harvard University Origins of Life Initiative (HUOLI), the Scottish Universities Physics Alliance (SUPA), the University of Geneva, the Smithsonian Astrophysical Observatory (SAO), and the Italian National Astrophysical Institute (INAF), the University of St Andrews, Queen's University Belfast, and the University of Edinburgh. 
This paper includes data collected by the \emph{Kepler}\ mission. Funding for the \emph{Kepler}\ mission is provided by the NASA Science Mission directorate. Some of the data presented in this paper were obtained from the Mikulski Archive for Space Telescopes (MAST). STScI is operated by the Association of Universities for Research in Astronomy, Inc., under NASA contract NAS5--26555. Support for MAST for non--HST data is provided by the NASA Office of Space Science via grant NNX13AC07G and by other grants and contracts.
This research has made use of NASA's Astrophysics Data System and the NASA Exoplanet Archive, which is operated by the California Institute of Technology, under contract with the National Aeronautics and Space Administration under the Exoplanet Exploration Program. 
This research has made use of the corner.py code by Dan Foreman-Mackey at \url{github.com/dfm/corner.py}.

Facilities: \facility{Kepler/K2}  \facility{TNG:HARPS-N}

\bibliographystyle{apj}
\bibliography{refs}
\clearpage

\appendix

\begin{figure}[]
\label{shark}
\centering
\includegraphics[scale=0.25]{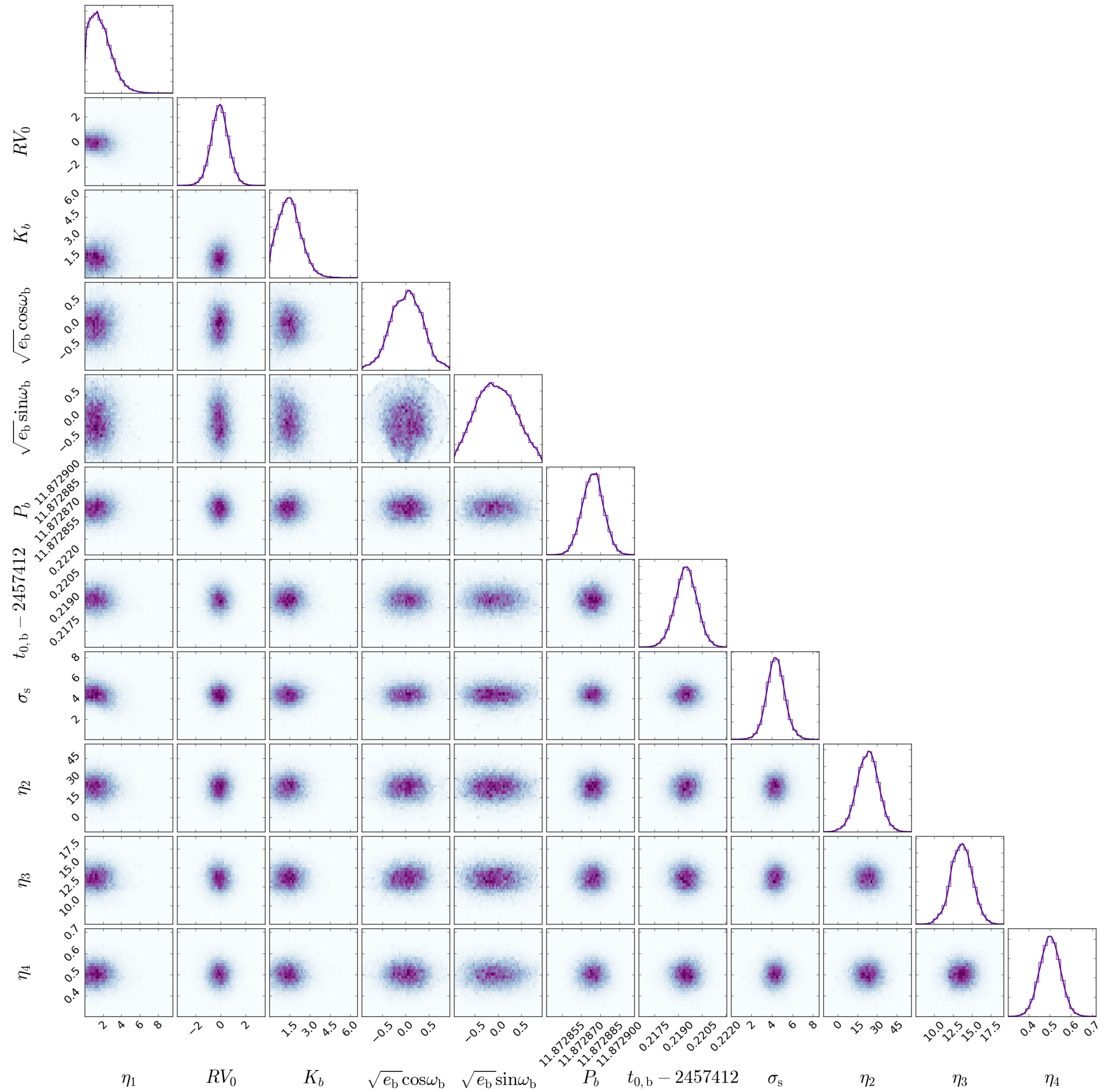}
\caption{Marginalized 1- and 2-D posterior distributions of the RV model parameters output from the MCMC procedure. 
The solid lines overplotted on the histograms are kernel density estimations of the marginal distributions.
The smooth, Gaussian-shaped posterior distributions attest of the good convergence of the MCMC chains.}
\end{figure}

\begin{table}[]
\centering
\caption{HARPS-N RV observations and spectroscopic activity indicators, determined from the DRS. From left to right are given: barycentric Julian date BJD, radial-velocity $RV$, the estimated 1-$\sigma$ uncertainty on the RV ($\sigma_{RV}$), the full width at half-maximum ($FWHM$), contrast and line bisector of ($BIS$) of the cross-correlation function (as defined in \citet{Queloz:2001vs}), the Ca II activity indicator $\log R'_{HK}$ and its 1-$\sigma$ uncertainty $\sigma_{\log R'_{HK}}$.}
\label{specdata}
\begin{tabular}{llllllll}
Barycentric Julian Date & \textit{RV} & \textit{$\sigma_{RV}$} & \textit{FWHM} & contrast & \textit{BIS} & $\log R'_{HK}$ & $\sigma_{\log R'_{HK}}$ \\

 [UTC] & [km.s$^{-1}$] & [km.s$^{-1}$] & [km.s$^{-1}$] & & [km.s$^{-1}$] &       &  \\
\hline
 \rule{0pt}{0ex} \\

2457180.523500 & -40.63968 & 0.00270 & 7.86870 & 29.321 & 0.02594 & -4.9614 & 0.0157 \\
2457181.527594 & -40.63651 & 0.00236 & 7.86380 & 29.310 & 0.02474 & -4.9651 & 0.0128 \\
2457182.603785 & -40.63892 & 0.00259 & 7.86953 & 29.305 & 0.02778 & -4.9600 & 0.0147 \\
2457183.494217 & -40.63709 & 0.00281 & 7.86066 & 29.369 & 0.04133 & -4.9601 & 0.0161 \\
2457184.498702 & -40.63199 & 0.00463 & 7.85989 & 29.256 & 0.03136 & -4.9596 & 0.0354 \\
2457185.495085 & -40.64455 & 0.00293 & 7.87730 & 29.309 & 0.03152 & -4.9532 & 0.0178 \\
2457186.572836 & -40.63822 & 0.00226 & 7.86260 & 29.343 & 0.02592 & -4.9750 & 0.0118 \\
2457188.501974 & -40.64006 & 0.00396 & 7.88033 & 29.250 & 0.03398 & -4.9270 & 0.0263 \\
2457189.492822 & -40.64273 & 0.00415 & 7.86784 & 29.305 & 0.03193 & -4.9742 & 0.0319 \\
2457190.506147 & -40.63649 & 0.00296 & 7.85754 & 29.330 & 0.01650 & -4.9878 & 0.0207 \\
2457191.506484 & -40.63570 & 0.00232 & 7.87102 & 29.344 & 0.02833 & -4.9906 & 0.0132 \\
2457192.503233 & -40.64261 & 0.00240 & 7.85594 & 29.332 & 0.03865 & -4.9884 & 0.0140 \\
2457193.506439 & -40.63887 & 0.00259 & 7.86547 & 29.344 & 0.02939 & -4.9654 & 0.0137 \\
2457195.618836 & -40.64015 & 0.00320 & 7.85577 & 29.331 & 0.02057 & -4.9571 & 0.0206 \\
2457221.430559 & -40.64291 & 0.00239 & 7.86913 & 29.361 & 0.02674 & -4.9743 & 0.0133 \\
2457222.435839 & -40.64084 & 0.00301 & 7.87471 & 29.314 & 0.02747 & -4.9653 & 0.0191 \\
2457223.460747 & -40.64063 & 0.00449 & 7.88822 & 29.216 & 0.03258 & -5.0076 & 0.0397 \\
2457224.390932 & -40.62888 & 0.00576 & 7.86308 & 29.208 & 0.04897 & -4.9174 & 0.0493 \\
2457225.433813 & -40.64082 & 0.00533 & 7.87820 & 29.198 & 0.02494 & -5.0425 & 0.0564 \\
2457226.408684 & -40.64007 & 0.00408 & 7.86647 & 29.256 & 0.01841 & -4.9618 & 0.0324 \\
2457227.450637 & -40.63983 & 0.00331 & 7.84180 & 29.309 & 0.03703 & -4.9530 & 0.0217 \\
2457228.410287 & -40.63623 & 0.00341 & 7.87077 & 29.293 & 0.02768 & -4.9568 & 0.0214 \\
2457229.429739 & -40.64052 & 0.00274 & 7.86533 & 29.298 & 0.03706 & -4.9567 & 0.0156 \\
2457230.406112 & -40.63739 & 0.00323 & 7.87646 & 29.312 & 0.02833 & -4.9308 & 0.0203 \\
2457254.397788 & -40.64424 & 0.00323 & 7.85502 & 29.314 & 0.02396 & -4.9604 & 0.0213 \\
2457255.500146 & -40.63523 & 0.00299 & 7.86901 & 29.308 & 0.02114 & -4.9755 & 0.0184 \\
2457256.421107 & -40.64243 & 0.00309 & 7.87620 & 29.285 & 0.02475 & -5.0159 & 0.0213 \\
2457257.482828 & -40.63653 & 0.00335 & 7.87228 & 29.271 & 0.01542 & -4.9738 & 0.0229 \\
2457267.507688 & -40.63665 & 0.00275 & 7.87107 & 29.284 & 0.02316 & -4.9523 & 0.0155 \\
2457268.565339 & -40.64117 & 0.00391 & 7.87216 & 29.272 & 0.02490 & -4.9535 & 0.0286 \\
2457269.463909 & -40.63953 & 0.00294 & 7.88436 & 29.297 & 0.02912 & -4.9824 & 0.0188 \\
2457270.452464 & -40.63594 & 0.00244 & 7.86921 & 29.341 & 0.03088 & -4.9645 & 0.0127 \\
2457271.453922 & -40.63895 & 0.00217 & 7.86716 & 29.344 & 0.02627 & -4.9705 & 0.0106 \\
2457272.495252 & -40.63976 & 0.00269 & 7.87068 & 29.289 & 0.03039 & -4.9893 & 0.0161 \\
2457273.471826 & -40.64459 & 0.00263 & 7.86265 & 29.320 & 0.03018 & -4.9924 & 0.0149 \\
2457301.432438 & -40.63609 & 0.00280 & 7.86113 & 29.327 & 0.01631 & -5.0003 & 0.0171 \\
2457302.432090 & -40.63881 & 0.00300 & 7.86872 & 29.290 & 0.03184 & -4.9895 & 0.0177 \\
2457322.359064 & -40.64060 & 0.00339 & 7.86850 & 29.200 & 0.03729 & -4.9534 & 0.0215 \\
2457324.381964 & -40.64046 & 0.00310 & 7.85083 & 29.271 & 0.03889 & -4.9723 & 0.0192 \\
2457330.349610 & -40.63902 & 0.00326 & 7.86853 & 29.288 & 0.01826 & -4.9757 & 0.0214 \\
2457331.372976 & -40.63873 & 0.00400 & 7.84861 & 29.288 & 0.03181 & -4.9975 & 0.0315 \\
2457332.370233 & -40.64343 & 0.00392 & 7.88092 & 29.280 & 0.03704 & -4.9701 & 0.0289 \\
2457333.372096 & -40.64721 & 0.00328 & 7.87162 & 29.263 & 0.02773 & -5.0009 & 0.0246 \\
2457334.328880 & -40.63632 & 0.00305 & 7.86380 & 29.304 & 0.02669 & -4.9949 & 0.0201 \\
2457336.372433 & -40.63546 & 0.00371 & 7.87532 & 29.294 & 0.02301 & -4.9793 & 0.0267 \\
2457498.664041 & -40.64669 & 0.00518 & 7.85834 & 29.142 & 0.00964 & -4.9501 & 0.0413 \\
2457499.669841 & -40.63127 & 0.00428 & 7.85774 & 29.202 & 0.02395 & -4.9950 & 0.0349 \\
2457521.623607 & -40.63859 & 0.00294 & 7.87430 & 29.328 & 0.03426 & -4.9826 & 0.0181 \\
2457522.593255 & -40.64256 & 0.00331 & 7.85636 & 29.278 & 0.03021 & -5.0042 & 0.0245 \\
2457525.643222 & -40.64070 & 0.00368 & 7.85232 & 29.286 & 0.04013 & -4.9688 & 0.0257 \\
2457526.666772 & -40.63541 & 0.00447 & 7.85574 & 29.270 & 0.02673 & -4.9869 & 0.0360 \\
2457527.615676 & -40.64328 & 0.00429 & 7.86272 & 29.214 & 0.03598 & -4.9852 & 0.0342 \\
2457528.632060 & -40.63341 & 0.00328 & 7.86301 & 29.249 & 0.02537 & -4.9359 & 0.0199 \\
2457529.644972 & -40.63372 & 0.00324 & 7.86321 & 29.255 & 0.02607 & -4.9483 & 0.0205 \\
2457530.649803 & -40.63592 & 0.00418 & 7.85874 & 29.177 & 0.02875 & -5.0283 & 0.0366 \\
2457531.672726 & -40.63028 & 0.00382 & 7.85133 & 29.244 & 0.01415 & -4.9776 & 0.0283 \\
2457557.640527 & -40.64505 & 0.00403 & 7.84956 & 29.207 & 0.03445 & -4.9784 & 0.0301 \\
2457558.611395 & -40.63639 & 0.00426 & 7.83640 & 29.255 & 0.03740 & -4.9370 & 0.0298 \\
2457559.640215 & -40.64116 & 0.00355 & 7.87551 & 29.164 & 0.02475 & -4.9554 & 0.0227 \\
2457560.634300 & -40.64283 & 0.00378 & 7.85731 & 29.167 & 0.02360 & -4.9598 & 0.0255 \\
2457562.593091 & -40.64033 & 0.00377 & 7.86254 & 29.244 & 0.02591 & -4.9733 & 0.0264 \\
2457563.620948 & -40.63391 & 0.00427 & 7.87441 & 29.158 & 0.04132 & -4.9638 & 0.0310 \\
2457564.607576 & -40.64620 & 0.00472 & 7.84112 & 29.226 & 0.02182 & -4.9581 & 0.0359 \\
2457565.628463 & -40.64556 & 0.00408 & 7.86775 & 29.249 & 0.02912 & -4.9645 & 0.0290 \\
2457566.630680 & -40.63824 & 0.00246 & 7.87116 & 29.285 & 0.01880 & -4.9889 & 0.0133 \\
2457573.575425 & -40.63182 & 0.00374 & 7.85929 & 29.305 & 0.02756 & -4.9720 & 0.0264 

\end{tabular}
\end{table}

\begin{table}[]
\contcaption{}
\centering

\begin{tabular}{llllllll}
Barycentric Julian Date & \textit{RV} & \textit{$\sigma_{RV}$} & \textit{FWHM} & contrast & \textit{BIS} & $\log R'_{HK}$ & $\sigma_{\log R'_{HK}}$ \\

 [UTC] & [km.s$^{-1}$] & [km.s$^{-1}$] & [km.s$^{-1}$] & & [km.s$^{-1}$] &       &  \\
\hline
 \rule{0pt}{0ex} \\
 2457573.597046 & -40.63485 & 0.00329 & 7.86519 & 29.315 & 0.02634 & -4.9879 & 0.0221 \\
2457574.566779 & -40.63346 & 0.00274 & 7.87878 & 29.312 & 0.02665 & -4.9477 & 0.0149 \\
2457574.586467 & -40.63648 & 0.00278 & 7.86857 & 29.326 & 0.02920 & -4.9623 & 0.0159 \\
2457576.557217 & -40.63311 & 0.00246 & 7.87205 & 29.311 & 0.02995 & -4.9681 & 0.0128 \\
2457579.629640 & -40.63536 & 0.00430 & 7.85924 & 29.250 & 0.02335 & -4.9734 & 0.0317 \\
2457580.702970 & -40.64565 & 0.00629 & 7.86485 & 29.182 & 0.02583 & -4.9331 & 0.0533 \\
2457602.492695 & -40.63907 & 0.00306 & 7.85957 & 29.311 & 0.03089 & -4.9285 & 0.0172 \\
2457614.470313 & -40.63970 & 0.00287 & 7.86161 & 29.292 & 0.03215 & -4.9681 & 0.0165 \\
2457616.518107 & -40.62758 & 0.00751 & 7.84425 & 29.065 & 0.03880 & -4.9471 & 0.0676 \\
2457617.485597 & -40.64913 & 0.00426 & 7.84666 & 29.192 & 0.03402 & -4.9775 & 0.0325 \\
2457618.483689 & -40.64341 & 0.00402 & 7.86390 & 29.124 & 0.02040 & -4.9969 & 0.0305 \\
2457651.405471 & -40.64647 & 0.00294 & 7.86796 & 29.302 & 0.02770 & -4.9582 & 0.0171 \\
2457652.404666 & -40.63602 & 0.00277 & 7.87473 & 29.349 & 0.02870 & -4.9674 & 0.0154 \\
2457653.410155 & -40.63573 & 0.00393 & 7.86262 & 29.305 & 0.02438 & -4.9947 & 0.0298 \\
2457654.408874 & -40.63807 & 0.00344 & 7.86736 & 29.345 & 0.02750 & -4.9921 & 0.0236 \\
2457655.380684 & -40.63649 & 0.00424 & 7.86868 & 29.252 & 0.02368 & -4.9915 & 0.0327 \\
2457656.400628 & -40.63716 & 0.00251 & 7.85571 & 29.329 & 0.02444 & -4.9617 & 0.0132 \\
2457658.467841 & -40.63825 & 0.00711 & 7.89902 & 29.261 & 0.02978 & -5.0566 & 0.0775 \\
2457659.409674 & -40.63750 & 0.00321 & 7.86655 & 29.335 & 0.02576 & -4.9494 & 0.0194 \\
2457661.433530 & -40.64007 & 0.00263 & 7.86886 & 29.389 & 0.02541 & -4.9687 & 0.0146 \\
2457669.401610 & -40.63981 & 0.00268 & 7.87027 & 29.322 & 0.03286 & -4.9664 & 0.0149 \\
2457670.357039 & -40.62378 & 0.00324 & 7.85478 & 29.319 & 0.02758 & -4.9473 & 0.0196 \\
2457671.396539 & -40.63589 & 0.00260 & 7.87140 & 29.334 & 0.03001 & -4.9790 & 0.0146 \\
2457672.400889 & -40.63547 & 0.00289 & 7.87136 & 29.313 & 0.02379 & -4.9728 & 0.0174 \\
2457673.332071 & -40.63531 & 0.00276 & 7.86742 & 29.302 & 0.02206 & -5.0001 & 0.0160 \\
2457699.366303 & -40.64003 & 0.00394 & 7.86373 & 29.291 & 0.03020 & -4.9620 & 0.0266 \\
2457701.363688 & -40.63710 & 0.00480 & 7.85277 & 29.252 & 0.04106 & -5.0048 & 0.0422 \\
2457702.373816 & -40.64049 & 0.00819 & 7.89468 & 29.016 & 0.03353 & -4.8835 & 0.0685 \\
2457706.345872 & -40.64216 & 0.00357 & 7.87407 & 29.227 & 0.04043 & -4.8960 & 0.0203

\end{tabular}
\end{table}
    

\end{document}